\documentclass[journal]{IEEEtran}
 \usepackage{caption}%
\usepackage{nomencl}
\usepackage{stackengine}
\newcommand\xrowht[2][0]{\addstackgap[.5\dimexpr#2\relax]{\vphantom{#1}}}
\usepackage{multicol}
\usepackage{multirow,array}
\usepackage{lipsum}
\usepackage{makecell, multirow}
\usepackage{mwe}
\usepackage{multirow}
\usepackage{graphicx}
\usepackage{color}
\usepackage{amsfonts}
\usepackage{float}
\usepackage{amsmath}
\usepackage{amsthm}
\usepackage{epstopdf}
\usepackage{changepage}
\usepackage{latexsym}
\usepackage{siunitx}
\usepackage{tikz}
\usetikzlibrary{arrows}
\usetikzlibrary{automata}
\usetikzlibrary{er}
\usetikzlibrary{positioning}
\usetikzlibrary{shapes,arrows}
\usetikzlibrary{calc}
\usepackage{relsize}
\usepackage{scalefnt}
\usepackage{caption}
\usepackage[margin=10pt,font=footnotesize,justification=centering]{subcaption}
\usepackage[margin=10pt,font=footnotesize,justification=centering,labelsep=period]{caption}
\usepackage{cases}
\usepackage{cite}
\usepackage{url}

\tikzset{fontscale/.style = {font=\relsize{#1}}}

\makeatletter
\newcommand{\vast}{\bBigg@{3.5}}
\newcommand{\Vast}{\bBigg@{5}}

\usepackage{mwe}
\usepackage{url}
\def\BibTeX{{\rm B\kern-.05em{\sc i\kern-.025em b}\kern-.08em
    T\kern-.1667em\lower.7ex\hbox{E}\kern-.125emX}}
    
\usepackage{stackengine}

\begin{document}

\title{Resilience Analysis and Cascading Failure Modeling of Power Systems under \\ Extreme Temperatures}
\author{Seyyed Rashid Khazeiynasab, ~\IEEEmembership{Student Member,~IEEE} and Junjian~Qi,~\IEEEmembership{Senior Member,~IEEE} 
\thanks{J. Qi's work was supported by National Science Foundation under Grant CAREER 1942206. 

S. R. Khazeiynasab is with the Department of Electrical and Computer Engineering, University of Central Florida, Orlando, FL 32816 USA (e-mail: rashid@knights.ucf.edu). 

J.~Qi (corresponding author) is with the Department of Electrical and Computer Engineering, Stevens Institute of Technology, Hoboken, NJ 07030 USA (e-mail: \mbox{jqi8@stevens.edu)}.

}
}

\maketitle

\begin{abstract}
In this paper, we propose an AC power flow based cascading failure model that explicitly considers external weather conditions, extreme temperatures in particular, and evaluates the impact of extreme temperature on the initiation and propagation of cascading blackouts. Specifically, load and dynamic line rating changes are modeled due to temperature disturbance, the probabilities for transmission line and generator outages are evaluated, and the timing for each type of events is carefully calculated to decide the actual event sequence. It should be emphasized that the correlated events, in the advent of external temperature changes, could together contribute to voltage instability. Besides, we model undervoltage load shedding and operator re-dispatch as control strategies for preventing the propagation of cascading failures. The effectiveness of the proposed model is verified by simulation results on the RTS-96 3-area system and it is found that temperature disturbances can lead to correlated load change and line/generator tripping, which together will greatly increase the risk of cascading and voltage instability. Critical temperature change, critical area with temperature disturbance, identification of most vulnerable buses, and comparison of different control strategies are also carefully investigated. 
\end{abstract}


\makenomenclature
\mbox{}

\nomenclature[A$$\Delta_a$$]{{$\Delta\lambda^d_{i,j}$}}{{Difference between the longitude of the buses $i$ and $j$ }} 
\nomenclature[A$$\Delta_b$$]{{$\Delta\phi^d_{i,j}$}}{{Difference between the latitude of buses $i$ and $j$ }} 

\nomenclature[A$$\Delta_c$$]{{$\Delta \mathrm{P}^{\mathrm{sh}}_i$}}{{The amount of active power to be shed at bus $i$}}

\nomenclature[A$$\Delta_c$$]{{$\Delta \mathrm{Q}^{\mathrm{sh}}_i$}}{{The amount of reactive power to be shed at bus $i$}}

\nomenclature[A$$\Delta_e $$]{{$\Delta o_{g,m}$}}{{ Accumulated overload in iteration $m$ (between $t_{m-1}$ and $t_{m}$) for generator $g$}}

\nomenclature[A$$\Delta_f$$]{{$\Delta o_{ij,m}$}}{{ Accumulated overload in iteration $m$ (between $t_{m-1}$ and $t_{m}$) for line $l:i \rightarrow j$ }}

\nomenclature[A$$\eta$$]{{$\eta$}}{{A constant which compensates the flow adjustment error due to the nonlinear nature of the power flow}}

\nomenclature[A$$\gamma$$]{{$\gamma$}}{{A constant, $0<\gamma\le 1$, where defined the size of the selected area}}

\nomenclature[A$$\lambda$$]{{$\overline{\lambda}$}}{{Upper bound
for longitude in selected area}}

\nomenclature[A$$\lambda$$]{{$\lambda_c$}}{{Latitude of the selected load bus}}

\nomenclature[A$$\lambda$$]{{$\underline{\lambda}$}}{{Lower bound for longitude in selected area}}
\nomenclature[A$$\lambda$$]{{$\lambda^{\min}$}}{{Minimum longitude among all buses}}

\nomenclature[A$$\lambda$$]{{$\lambda^{\max}$}}{{Maximum longitude among all buses}}

\nomenclature[A$$\phi$$]{{$\phi_c$}}{{Longitude of the selected load bus}}

\nomenclature[A$$\phi$$]{{$\phi^{\max}$}}{{Maximum latitude among all buses}}

\nomenclature[A$$\phi$$]{{$\phi^{\min}$}}{{Minimum latitude among all buses}}

\nomenclature[A$$\phi_y$$]{{$\overline{\phi}$}}{{Upper bound for latitude in a selected area}}

\nomenclature[A$$\phi_z$$]{{$\underline{\phi}$}}{{Lower bound for latitude in a selected area}}

\nomenclature[A$D$]{{$D_{ij}$}}{{Distance between bus $i$ and $j$}}
 
\nomenclature[A$F$]{{$\overline{F}_{ij}^0$}}{{Initial rating of line $l:i \rightarrow j$ }}

\nomenclature[A$F$]{{$F_{ij}$}}{{Flow of line $l:i \rightarrow j$ }}

\nomenclature[A$F$]{{$\overline{F}^\mathrm{d}_{ij}$}}{{Dynamic line rating of line $l:i \rightarrow j$ }}

\nomenclature[A$o$]{{$\bar{o}_{ij}$}}{{Overload limit of line $l:i \rightarrow j$ }}

\nomenclature[A$o$]{{$\bar{o}_g$}}{{Overload limit of generator $g$ }}

\nomenclature[A$K$]{{$k_\mathrm{sh}$}}{{Load shedding constant}}

\nomenclature[A$P$]{{$\overline{\mathrm{P}}_i$}}{{Active power capacity of generator $i$}}

\nomenclature[A$P$]{{$\mathrm{P}_{P_i}$}}{{Real power output of generator $\mathrm{G}_{P_i}$ after re-dispatch}} 

\nomenclature[A$P$]{{$\mathrm{P}^0_{P_i}$}}{{Real power output of generator $\mathrm{G}_{P_i}$ before re-dispatch}}

\nomenclature[A$P$]{{$\mathrm{P}_i^0$}}{{Nominal real power at bus $i$}}

\nomenclature[A$P$]{{$P^\mathrm{trip}_{ij}$}}{{The tripping probability of the line  $l:i \rightarrow j$}}

\nomenclature[A$P$]{{$P^\mathrm{trip}_{i}$}}{{The tripping probability of the generator $i$}}

\nomenclature[A$P$]{{$\mathrm{P}_i$}}{{The real power of load bus $i$}}

\nomenclature[A$P$]{{$\mathrm{pf}_i$}}{{The power factor of bus $i$}}

\nomenclature[A$T$]{{$T_{ij}^0$}}{{The initial temperature of line $l:i \rightarrow j$}}

\nomenclature[A$P$]{{$\mathrm{pf}_i^0$}}{{Initial power factor under initial temperature at bus $i$}}

\nomenclature[A$Q$]{{$\underline{\mathrm{Q}}_i$}}{{Lower reactive power capacity for generator $i$}}

\nomenclature[A$Q$]{{$\overline{\mathrm{Q}}_i$}}{{Upper reactive power capacity for generator $i$}}

\nomenclature[A$R$]{{$R$}}{{Radius of the earth}}

\nomenclature[A$S$]{{$\mathcal{S}_\mathrm{B}$}}{{Set of buses}}

\nomenclature[A$S$]{{$\mathcal{S}_\mathrm{L}$}}{{Set of load buses}}

\nomenclature[A$S$]{{$\mathcal{S}_\mathrm{L}^A$}}{{Set of load buses within the selected area}}

\nomenclature[A$S$]{{$\mathcal{S}_\mathrm{G}$}}{{Set of generator buses}}

\nomenclature[A$S$]{{$\mathcal{S}_\mathrm{line}$}}{{ Set of transmission lines}}

\nomenclature[A$T$]{{$T^0$}}{{Initial ambient temperature for all load bus}}

\nomenclature[A$T$]{{$T_i$}}{{Ambient temperature where bus $i$ is located}}

\nomenclature[A$V$]{{$\mathrm{V}_{ij}^\mathrm{rated}$}}{{Nominal voltage of line $l:i \rightarrow j$}}
\nomenclature[A$V$]{{$\mathrm{V}_{ij}$}}{{Per unit voltage of line $l:i \rightarrow j$}}

\nomenclature[A$V$]{{$\mathrm{V}_{ij}^0$}}{{Initial per unit voltage of line $l:i \rightarrow j$}}


\section{Introduction} \label{sec:introduction}

\IEEEPARstart{C}{ascading} {failure is a common phenomenon in both natural and engineered systems, such as electric power systems \cite{sun2019power}, natural gas systems \cite{praks2017monte},  transportation networks  \cite{theoharidou2011securing}, disease transmission networks \cite{hong2016epidemic}, and interdependent networks \cite{hong2016cascading,hong2015failure, hong2017cascading}. }
For example, there have been several large-scale blackouts, such as  
the 2003 U.S.-Canadian blackout \cite{us2004final}, 
the 2011 Arizona-Southern California blackout \cite{11blackout}, and the 2012 Indian blackout \cite{indian}, which have led to extensive outage propagations and significant impacts \cite{haggi2019review}.  
In order to simulate and analyze cascading failures, many models with different levels of details have been developed \cite{sun2019power}, such as Manchester model \cite{Kirschen}, hidden failure model \cite{Phadke}, \cite{chen}, CASCADE model \cite{cascade}, OPA model \cite{opa4}, AC OPA model \cite{ACOPA1}, dynamic model \cite{dynamic}, sandpile model \cite{sandpile}, branching process model \cite{qi2013towards,dobson2012estimating}, multi-type branching process model \cite{qi2016estimating},  the interaction model \cite{qi2014interaction,ju2015simulation,qi2017efficient,Ju2017}, and Markovian influence graph \cite{zhou2020markovian}.

However, the existing models mainly focus on the system itself, usually ignoring the interactions between the system and various external factors, such as extreme weather conditions. These factors are important for both initiating and propagation of cascading failures. For U.S.-Canadian blackout on August 14, 2003, the temperature was high ($31^{\circ}\mathrm{C}$), causing load increase in FirstEnergy's control area, transmission line tripping due to tree contact, and generator tripping due to increased reactive power outputs \cite{us2004final}. 
Another blackout occurred partly because of temperature disturbance on July 2, 1996. High loads in Southern Idaho and Utah due to high temperature (around  $\mathrm{38^{\circ}\mathrm{C}}$)\cite{kosterev1999model,taylor1997recording} led to high demands and subsequently highly loaded transmission lines. 

In recent years, a few papers have investigated the impact of temperature on cascading failure risks. 
In \cite{anghel2007stochastic}, a stochastic model is proposed in which random line failures are generated at constant failure rates and  overloaded-line failures occur when the line temperature reaches the equilibrium temperature. In \cite{slow}, an OPA model with slow process is proposed in which the line temperature evolution is modeled for calculating the line length and sag changes in order to evaluate the possibility for tree contact or damage. In \cite{pra}, a PRA model is developed to consider the impact of wind speed and the evolution of line temperature. In \cite{yao2018towards}, risk assessment of weather-related cascading outages is presented based on weather-dependent outage rates. In \cite{dobson2017exploring},  historical outage data is used to estimate the effects of weather on cascading failure and bulk statistics of historical initial line outages are provided. However, all these existing models have the following problems. 
\begin{enumerate}
	\item The initiating events are still generated by random sampling, which does not consider the important geographical correlations of the initiating events due to external weather conditions such as temperature disturbance. 
	\item The ambient temperature disturbance at various geographical locations in the system is not explicitly modeled, and the consequent demand changes and dynamic rating changes are not modeled, which, however, are very critical for understanding the initiating and propagation of cascading failures especially due to loss of voltage stability. 
\end{enumerate}

Therefore, in order to better understand cascading failure it is needed to develop a model that could explicitly consider the ambient temperature disturbances and their impacts on system operation and cascading failure risks. The contributions of this paper are listed below.
\begin{enumerate}
	\item We develop a cascading failure model that explicitly considers ambient temperature disturbances and the subsequent demand change and dynamic line rating changes to take into account the correlations between different events such as line outage, generator tripping, and undervoltage of load buses, and models the control strategies against failure propagation such as undervoltage load shedding and operator re-dispatch. 
	\item Based on the developed cascading failure model we provide an explanation about why the failure can still be initiated and propagated even when the power system is initially $N-1$ secure by considering the impact of ambient temperature disturbances and the correlations between different events. 
	\item We perform risk assessment for power systems based on the developed model to investigate critical temperature change and critical area with temperature disturbance that could lead to significantly increased risk of cascading, identify the most vulnerable buses for temperature disturbances, and evaluate the effectiveness of different control strategies on reducing the system risk. 
\end{enumerate}

The remainder of this paper is organized as follows. Section \ref{ambient_temp} discusses ambient temperature disturbance, models load and line rating changes due to temperature change, and evaluates the probability of transmission line and generator tripping. Control strategies including undervoltage load shedding and operator re-dispatch are modeled in Section \ref{control_sec}. Section \ref{event_time} determines the timing of different types of events, and Section \ref{vsm_intro} introduces voltage stability margin calculation. In Section \ref{model-summary} the proposed blackout model is summarized. Section \ref{result} tests and validates the proposed model on the RTS-96 3-area system. Finally, conclusions are drawn in Section \ref{conclusion}.

\section{Ambient Temperature in Blackout Modeling}\label{ambient_temp}

Assume there are $n$ buses in a power system, including a slack bus that is numbered as $i_\mathrm{s}$. 
The vector of ambient temperatures of all buses is $\boldsymbol{T}=[T_1,T_2,\cdots,T_n]^\top$. 
The vector of ambient temperatures of the load buses is denoted by $\boldsymbol{T}_{\mathcal{S}_\mathrm{L}}$. 
For a transmission line $l:i \rightarrow j$ that connects bus $i$ and bus $j$ and crosses $M$ areas, its ambient temperature is assumed to be dependent on the temperatures of the $M$ areas.

\subsection{Temperature Disturbance} \label{tem_dist}

An ambient temperature disturbance is applied to an area $A=[\underline{\phi},\overline{\phi}] \times [\underline{\lambda},\overline{\lambda}]$.
In order to make sure at least one load bus is inside the chosen area, we randomly select one of the load buses and choose an area around this bus. The chosen area is set as 
$\underline{\phi}=\phi_c-\Delta \phi$, $\overline{\phi}=\phi_c+\Delta \phi$,  $\underline{\lambda}=\lambda_c-\Delta \lambda$, and $\overline{\lambda}=\lambda_c+\Delta \lambda$ where $\Delta \phi>0$ and $\Delta \lambda>0$ determine how widespread the disturbance is and are chosen as 
\begin{align}
&\Delta \phi = \gamma (\phi^{\max}-\phi^{\min}) \\
&\Delta \lambda = \gamma (\lambda^{\max}-\lambda^{\min}).
\end{align}
As a disturbance the ambient temperature of the load buses in the selected subsystem is changed by $\Delta T$. 
Obviously the ambient temperature of the transmission lines in the selected subsystem will also change by $\Delta T$. 
For a line $l:i \rightarrow j$ that crosses the boundary of the selected area and lies in $M$ areas, its length in area $k$ with ambient temperature of $T_\mathrm{k}$ is $d_k$ and its temperature is determined by 
\begin{align}
T_{ij} = \frac{1}{D_{ij}}\sum_{k=1}^{M} T_k d_k,
\end{align}
where bus $i$ is assumed to be inside the selected area while bus $j$ is not and $D_{ij}$ is the distance between buses $i$ and $j$. As a simple case, for a line $l:i \rightarrow j$ that only crosses two areas, its temperature is given as
 \begin{align}
T_{ij} = \frac{d_1}{D_{ij}}T_i + \frac{d_2}{D_{ij}}T_j,
\end{align} 
where $d_1$ is the length of the line in the selected area and $d_2$ is the length of the line out of the selected area. Then with a $\Delta T$ change for bus $i$ the ambient temperature of the line will change by 
\begin{align}
\Delta T_{ij}=\frac{d_{1}}{D_{ij}}\Delta T.
\end{align}
For calculating the distance between two buses, with specific latitude and longitude, we assumes that the earth is a sphere with a radius of 6378 km. We calculate the distance by using haversine formula in (\ref{haves1})--(\ref{haveslast}) \cite{rick1999deriving}. Let the central angle $\Theta$ between two buses $i$ and $j$ be: 
\begin{align}\label{haves1}
&\Theta=\frac{D_{ij}}{R}.
\end{align}
The haversine formula ($\mathrm{hav}$ of $\Theta$) is:
\begin{align}
\mathrm{hav}(\Theta)=\mathrm{hav}(\Delta\phi_{i,j}^d)+\cos\phi_i \cos\phi_j\,\mathrm{hav}(\Delta\lambda^d_{i,j}),
\end{align}
and the haversine function of an angle $\theta$ is:
\begin{align}
\mathrm{hav}(\theta)=\sin^2\bigg(\frac{\theta}{2}\bigg).
\end{align}
Finally, by applying the inverse haversine $\mathrm{hav}^{-1}$ to the central angle $\Theta$, we can find the distance $D_{ij}$ as \\
\begin{align}\label{haveslast}
D_{ij}=2\,R\,\sin^{-1}\Bigg(\sqrt{ \sin^2\frac{\Delta\phi^d_{i,j}}{2}+\cos\phi_i \cos\phi_j\sin^2\frac{\Delta\lambda^d_{i,j}}{2}}\Bigg). \notag
\end{align}

Note that a relatively large ambient temperature change could take a few hours during which some protections may operate. However, from the modeling perspective it may be too complicated if we  consider the temporal behavior of the ambient temperature change and its impact on the risk of cascading failures. Therefore, in this paper we consider a simplified scenario in which the ambient temperature change happens immediately and we focus more on what impact it will have on system operation and  cascading failure risks.

\subsection{Load Change under Temperature Disturbances} \label{load_temp_subsec}

In power systems, load forecasting is used for day-ahead generation purchases and reactive power management. 
However, due to uncertainties the actual load may be different from the forecasted load. 
For example, several large operators in the Midwest consistently under-forecasted the load levels between August 11 and 14, 2003 \cite{us2004final}. 
In this paper we assume $T^0$ is used for day-ahead load forecasting while the actual ambient temperature for a subsystem is $T^0+\Delta T$ which will lead to a deviation of actual load from forecasted load. 

The real power of a load bus $i$ changes with its ambient temperature $T_i$ as
\begin{align}
&\mathrm{P}_i = L_{\mathrm{P}_i}(T_i) \mathrm{P}_i^0. 
\end{align}
For simplicity and without loss of generality, we assume that $L_{\mathrm{P}_i}(T_i)=L(T_i)$ where the same function $L$ is used for all buses. 

According to \cite{temerature1,temerature2,temerature3}, $L(T_i)$ can be represented by a polynomial function as $L(T_i)=a_3 {T_i}^3 + a_2 {T_i}^2 + a_1 {T_i} + a_0$. 

\begin{figure}[!t]
	\captionsetup{justification=raggedright,singlelinecheck=false}
	\centering
	\centerline{\includegraphics[width=3in]{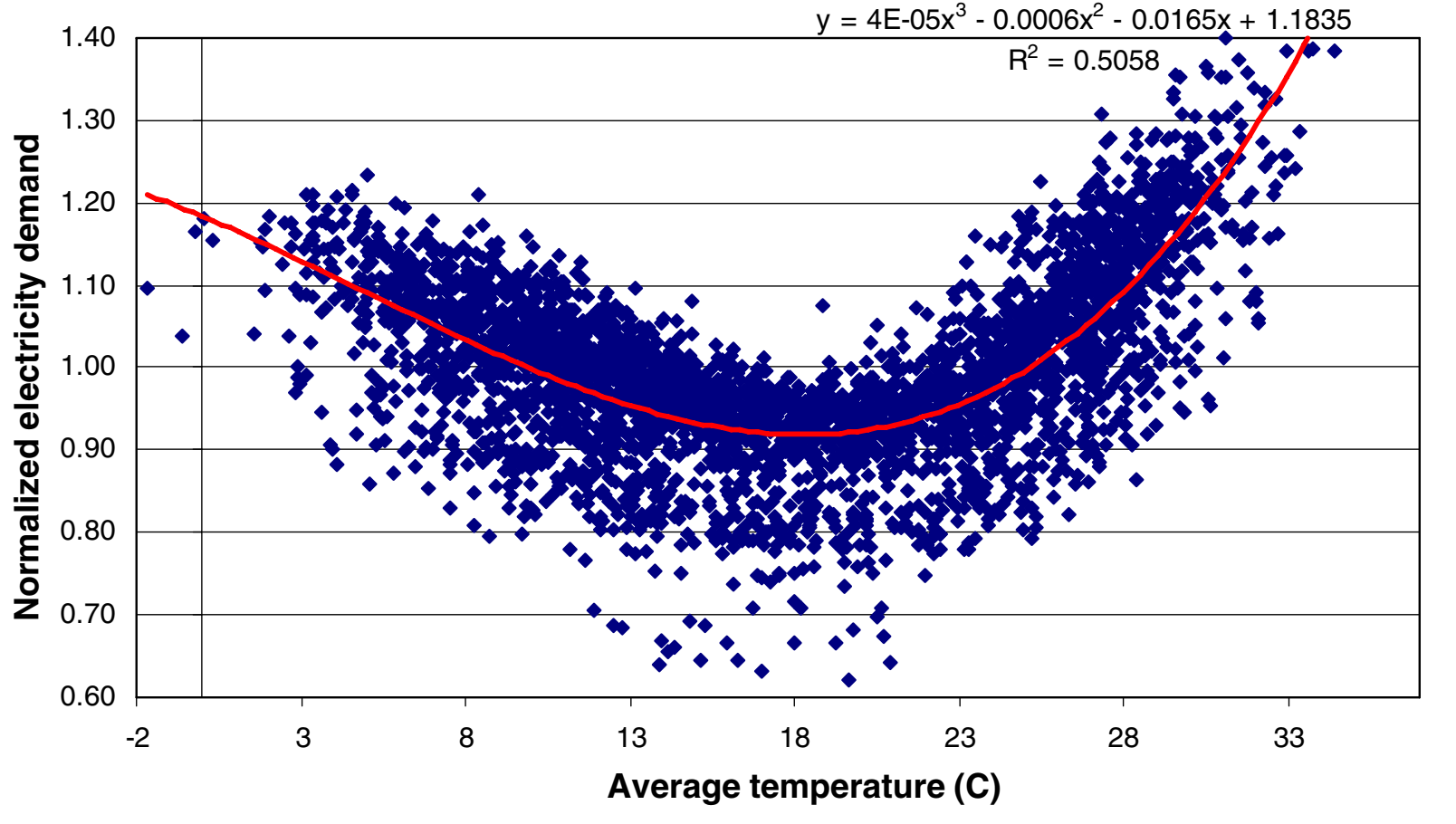}}
	\caption{Relationship between normalized electricity demand and temperature for the Greek power system based on daily data for the period 1/1/1993--12/31/2003 \cite{temerature2,temerature3}.}
	\label{load_temp}
\end{figure}

Fig. \ref{load_temp} shows the fitted $L$ function for the Greek interconnected power system based on data between January 1, 1993 and December 31, 2003 \cite{temerature2,temerature3}. It is seen that the variation of load with temperature is nonlinear and asymmetrically 
increasing for decreased or increased temperatures 
with a minimum at around $T_{\min}=18.5^{\circ}\mathrm{C}$ \cite{temerature2,temerature3}. 
The load is more sensitive to higher temperature increase than to lower temperature decrease, mainly because several energy sources such as diesel, natural gas, electricity can be used for heating while practically only electricity can be used for cooling \cite{temerature2,temerature3}. 
A similar curve can be found in \cite{temerature1}. 
Assume the initial load to be the ambient temperature that leads to $L(T)=1$. 
For example, if we use the curve in Fig. \ref{load_temp} there will be two corresponding positive ambient temperatures, which are $T_{\mathrm{low}}^0=9.91^{\circ}\mathrm{C}$ and $T_{\mathrm{high}}^0=24.21^{\circ}\mathrm{C}$, respectively. If we want to explore the effect of temperature increase (decrease), we will assume that the ambient temperature corresponding to the initial load is $T^0=T_{\mathrm{high}}^0$ ($T^0=T_{\mathrm{low}}^0$).

Under high/low temperatures, there will be more air conditioning loads which consume more reactive power and have lower power factors than other types of load \cite{us2004final}. In order to take this into account, we assume that the power factor linearly decreases with the temperature increase as
\begin{align}
\mathrm{pf}_i
= &\begin{cases}
\mathrm{pf}_i^0 - k_i^{\mathrm{pf}} (T_i-T^0), &\text{if $\, T^0=T_{\mathrm{high}}^0, T_i\ge T_\mathrm{min}$} \\
\mathrm{pf}_i^0 + k_i^{\mathrm{pf}} (T_i-T^0), &\text{if $\,T^0=T_{\mathrm{low}}^0, T_i \le T_\mathrm{min}$}, \notag
\end{cases}
\end{align}
Then the reactive power of load bus $i$ under temperature $T_i$ can be obtained as:
\begin{align}
\mathrm{Q}_i=\mathrm{P}_i \tan \big(\cos^{-1}(\mathrm{pf}_i)\big).
\end{align}
\subsection{Dynamic Line Rating} \label{dynamic_rating_sec}

Assume the initial rating of a transmission line $l:i \rightarrow j$, 
is determined for $T_{ij}^0=(T_i^0+T_j^0)/2$. When we study the scenario for high temperature, $T_{ij}^0=T_{\mathrm{high}}^0=24.21^{\circ}\mathrm{C}$. 
This is consistent with the fact that the approximate current carrying capacity is usually given for $25^{\circ}\mathrm{C}$ \cite{analysis}. 

The dynamic rating of a line can depend on ambient temperature \cite{rating} and utility management vegetation \cite{slow}. 
According to \cite{rating}, the effect of ambient temperature on dynamic rating expressed in Ampere is quasi linear. 
Therefore, for dynamic rating expressed in apparent power (MVA) we have
\begin{align} \label{dynamic_rating_formula0}
\overline{F}^\mathrm{d}_{ij} = \mathrm{V}_{ij}^\mathrm{rated}\mathrm{V}_{ij}(-k_{ij} T_{ij} + c_{ij}).
\end{align}
Different conductors may have different slopes $k_{ij}$. For example, the slope for the AMS570 conductor is approximately $0.02 \,\mathrm{kA}/^{\circ}\mathrm{C}$ \cite{rating}. For simplicity, in this paper we use the same slope $k_{ij}=0.02 \,\mathrm{kA}/^{\circ}\mathrm{C}$ for all lines. Then $c_{ij}$ can be easily obtained as $c_{ij}=\overline{F}^0_{ij}/(\mathrm{V}_{ij}^\mathrm{rated}\mathrm{V}_{ij}^0) + k_{ij} T^0_{ij}$.

In order to consider utility vegetation management and the corresponding risk for line tripping due to a slow process involving transmission line temperature evolution, sag increase, and tree contact \cite{slow}, we modify (\ref{dynamic_rating_formula0}) to be
\begin{align} \label{dynamic_rating_formula1}
\overline{F}_{ij}^\mathrm{d}=\alpha_{ij}\mathrm{V}_{ij}^\mathrm{rated}\mathrm{V}_{ij}(-k_{ij} T_{ij} + c_{ij}),
\end{align}
where $\alpha_{ij}$ is uniformly sampled in $[\underline{\alpha}, 1]$ with $0<\underline{\alpha}\le 1$. 
When $\alpha_{ij}=1$, the dynamic rating is only determined by ambient temperature. Otherwise, the dynamic rating will decrease. 

\subsection{Probability of Line Tripping} \label{line_prob_sec}

For a line $l:i \rightarrow j$, let $R_{ij}(\boldsymbol{T})=F_{ij}(\boldsymbol{T}_{\mathcal{S}_\mathrm{L}})/\overline{F}_{ij}^\mathrm{d}(T_{ij})$. Note that $F_{ij}$ is a function of the ambient temperatures of the load buses and will change when there is a load change at any load bus due to ambient temperature change. By contrast, the dynamic line rating $\overline{F}^\mathrm{d}_{ij}$ is only a function of the local ambient temperature of the line $l:i \rightarrow j$. 

\begin{figure}[!t]
\captionsetup{justification=raggedright,singlelinecheck=false}
\centering
\centerline{\includegraphics[width=0.70\columnwidth]{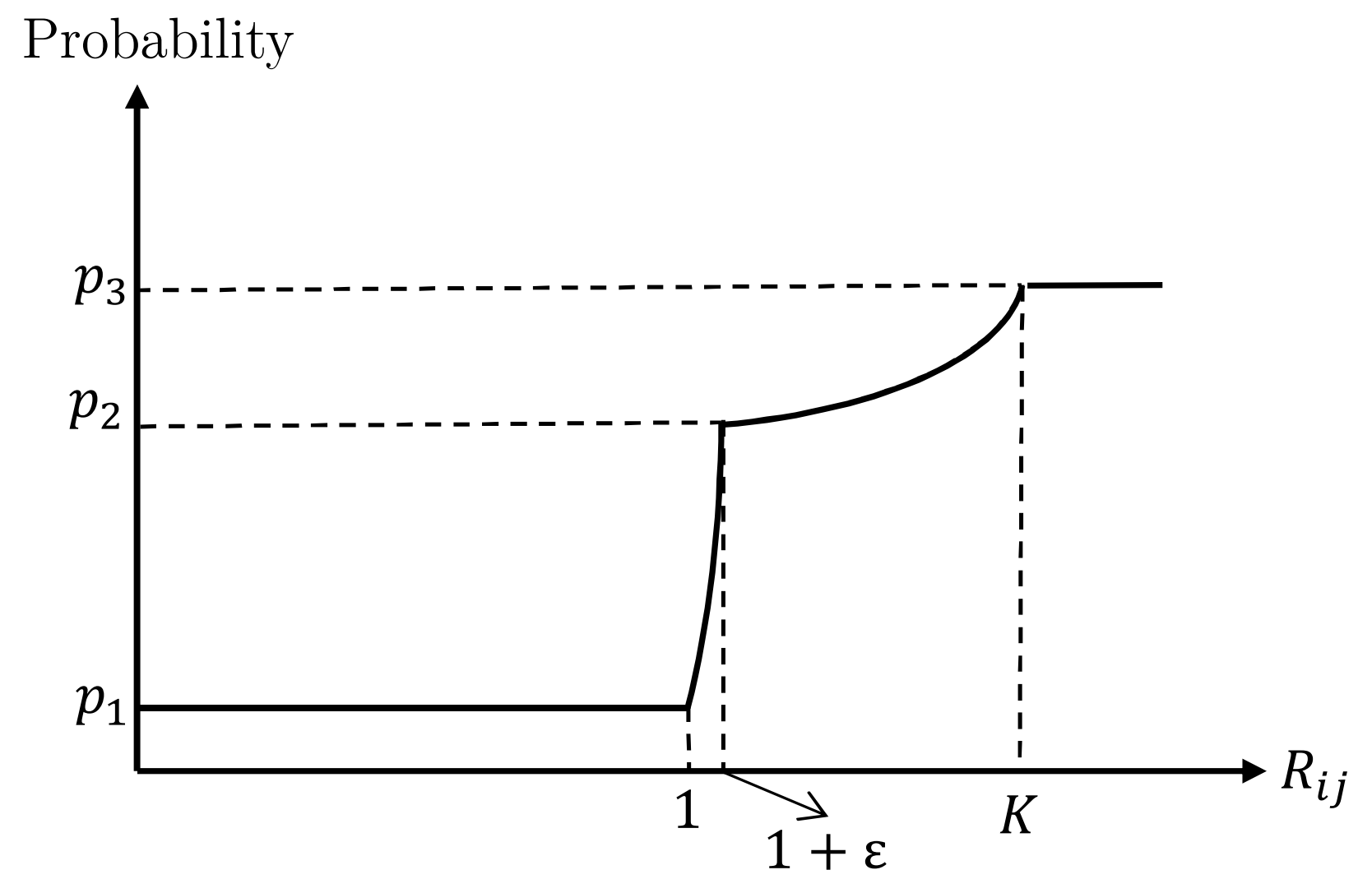}}
\caption{Probability of line tripping as a function of $R_{ij}$.}
\label{exponential}
\end{figure}

The tripping probability of the line can be written as a function of $R_{ij}$: 
\begin{align}
P^\mathrm{trip}_{ij} = f_\mathrm{t}\big(R_{ij}(\boldsymbol{T})\big).
\end{align}
In this paper we propose the function shown in Fig. \ref{exponential} for tripping probability of the line, which can be written as
\begin{align}
f_\mathrm{t}&= \begin{cases}
p_1, &\text{if $\, R_{ij} \le 1$} \\
a_1\kern 0.08em e^{b_1 R_{ij}}, &\text{if $\, 1 <R_{ij} \le 1+\epsilon$} \\
a_2\kern 0.08em e^{b_2R_{ij}}, &\text{if $\, 1+\epsilon < R_{ij} \le K$} \\
p_3, & \text{if $\,\,\, R_{ij} > K$},
\end{cases}
\end{align}
where $b_1=(\ln p_1 - \ln p_2)/(-\epsilon)$, $a_1=p_1/e^{b_1}$, $b_2=(\ln p_2 - \ln p_3)/(1+\epsilon-K)$, and $a_2=p_2/e^{b_2 (1+\epsilon)}$. When $R_{ij}\leq1$, although there is no slow process involved  a line may still be tripped by a small probability, by which we can take into account the factors that could lead to line tripping even if the dynamic line rating is not reached. For example, even if $F_{ij}$ is far below $F^d_{ij}$ the line may be tripped by the protection due to lighting strikes, then the probability of line tripping is not equal zero, but is equal to a constant low value  to describe a probability of the tripping of a line exposed to a hidden failure. 
Once $R_{ij}>1$ it becomes possible for the line to be tripped such as due to tree contact or overheating and the probability of tripping thus quickly grows to a much higher value $p_2$ at $R_{ij}=1+\epsilon$. Then between $1+\epsilon$ and $K$ the probability of tripping increases exponentially with parameters $a_2$ and $b_2$. Finally, it reaches to a high probability $p_3\le 1$ when $R_{ij}$ is greater than $K$. In this paper we choose $p_3$ to be 1.

\subsection{Probability of Generator Tripping}\label{probab_gen_tripping}

The active power change due to the ambient temperature change is $\Delta \mathrm{P}=\sum\limits_{i\in \mathcal{S}_\mathrm{L}^A}(\mathrm{P}_i-\mathrm{P}_i^0)$. 
This change will be supplied by all generators in proportion to their active power reserve. Specifically, for $i\in \mathcal{S}_\mathrm{G} \backslash i_\mathrm{s}$ we have
\begin{align} \label{P_change}
\mathrm{P}_i=\mathrm{P}_i^0+\frac{\overline{\mathrm{P}}_i-\mathrm{P}_i^0}{\sum\limits_{i\in \mathcal{S}_\mathrm{G}}(\overline{\mathrm{P}}_i-\mathrm{P}_i^0)}\Delta \mathrm{P}.
\end{align}

When reactive load is increased, the generators nearby will have to provide more reactive power. 
For example, in 2003 U.S.-Canadian blackout, Eastlake unit 5 in FirstEnergy's Northern Ohio service area was generating high reactive power, because there were significant reactive power supply problems in the states of Indiana and Ohio. Due to high reactive output and over-excitation, this unit was tripped \cite{us2004final}. 

Automatic voltage regulator is assumed to be equipped for each generator to hold the terminal voltages. 
Since normally there is no automatic control action limiting the reactive power output of generators \cite{kunder}, the reactive power of a generator can be beyond the allowed capacity range $[\underline{\mathrm{Q}}_i,\overline{\mathrm{Q}}_i]$ due to voltage regulation to relieve the overvoltage or undervoltage violations close to the generator. In this paper we consider the increased possibility of generator tripping due to overexcitation limiter operation. 
The tripping probability of the generator $i$ can be written as a function of $\mathrm{Q}_{i}$ as: 
\begin{align}
P^\mathrm{trip}_{i} = f_\mathrm{g}\big(\mathrm{Q}_{i}\big).
\end{align}
In this paper we propose the following function: 
\begin{align}
f_\mathrm{g}&= \begin{cases}
p_6, &\text{if $\, \mathrm{Q}_{i} \le \underline{K}_{\mathrm{Q}_i}$} \\
a_3\kern 0.08em e^{b_3 \mathrm{Q}_{i}}, &\text{if $\, \underline{K}_{\mathrm{Q}_i} < \mathrm{Q}_{i} \le \underline{\mathrm{Q}}_i-\underline{\epsilon} $} \\
a_4\kern 0.08em e^{b_4 \mathrm{Q}_{i}}, &\text{if $\, \underline{\mathrm{Q}}_i-\underline{\epsilon} < \mathrm{Q}_{i} \le \underline{\mathrm{Q}}_i $} \\
p_4, &\text{if $\,\underline{\mathrm{Q}}_i < \mathrm{Q}_{i} \le \overline{\mathrm{Q}}_i$} \\
a_5\kern 0.08em e^{b_5 \mathrm{Q}_{i}}, &\text{if $\, \overline{\mathrm{Q}}_i < \mathrm{Q}_{i} \le \overline{\mathrm{Q}}_i+\overline{\epsilon}$} \\
a_6\kern 0.08em e^{b_6 \mathrm{Q}_{i}}, &\text{if $\, \overline{\mathrm{Q}}_i+\overline{\epsilon} < \mathrm{Q}_{i} \le \overline{K}_{\mathrm{Q}_i}$} \\
p_6, & \text{if $\, \mathrm{Q}_{i} > \overline{K}_{\mathrm{Q}_i}$},
\end{cases}
\end{align}
where $b_3=(\ln p_6 - \ln p_5)/(\underline{K}_{\mathrm{Q}_i}-\underline{\mathrm{Q}}_i+\underline{\epsilon})$, $a_3=p_6/e^{b_3 \underline{K}_{\mathrm{Q}_i}}$, $b_4=(\ln p_5 - \ln p_4)/(-\underline{\epsilon})$, $a_4=p_5/e^{b_4 (\underline{\mathrm{Q}}_i-\underline{\epsilon})}$, $b_5=(\ln p_4 - \ln p_5)/(-\overline{\epsilon})$, $a_5=p_4/e^{b_5 \overline{\mathrm{Q}}_i}$, $b_6=(\ln p_5 - \ln p_6)/(\overline{\mathrm{Q}}_i+\overline{\epsilon}-\overline{K}_{\mathrm{Q}_i})$, and $a_6=p_5/e^{b_6 (\overline{\mathrm{Q}}_i+\overline{\epsilon})}$. 
If the reactive power of any generator $i$ lies in $[\underline{\mathrm{Q}}_i,\overline{\mathrm{Q}}_i]$, it fails only by a very small probability $p_4$ in order to model any accidental failure. When $\mathrm{Q}_i$ falls out of $[\underline{\mathrm{Q}}_i,\overline{\mathrm{Q}}_i]$, the probability for generator tripping quickly grows to a much higher value $p_5$ at $\underline{\mathrm{Q}}_i-\underline{\epsilon}$ or $\overline{\mathrm{Q}}_i+\overline{\epsilon}$. Then from $\underline{\mathrm{Q}}_i-\underline{\epsilon}$  to  $\underline{K}_{\mathrm{Q}_i}$ or from $\overline{\mathrm{Q}}_i+\overline{\epsilon}$  to  $\overline{K}_{\mathrm{Q}_i}$, the probability of tripping increases exponentially with parameters $a_3$, $b_3$ and $a_6$, $b_6$, respectively. Finally when $\mathrm{Q}_i \le \underline{K}_{\mathrm{Q}_i}$ or $\mathrm{Q}_i > \overline{K}_{\mathrm{Q}_i}$ it reaches a high probability $p_6$ for the most abnormal cases.

\begin{figure}[!t]
\centering
\captionsetup{justification=raggedright,singlelinecheck=false}
\centerline{\includegraphics[width=0.70\columnwidth]{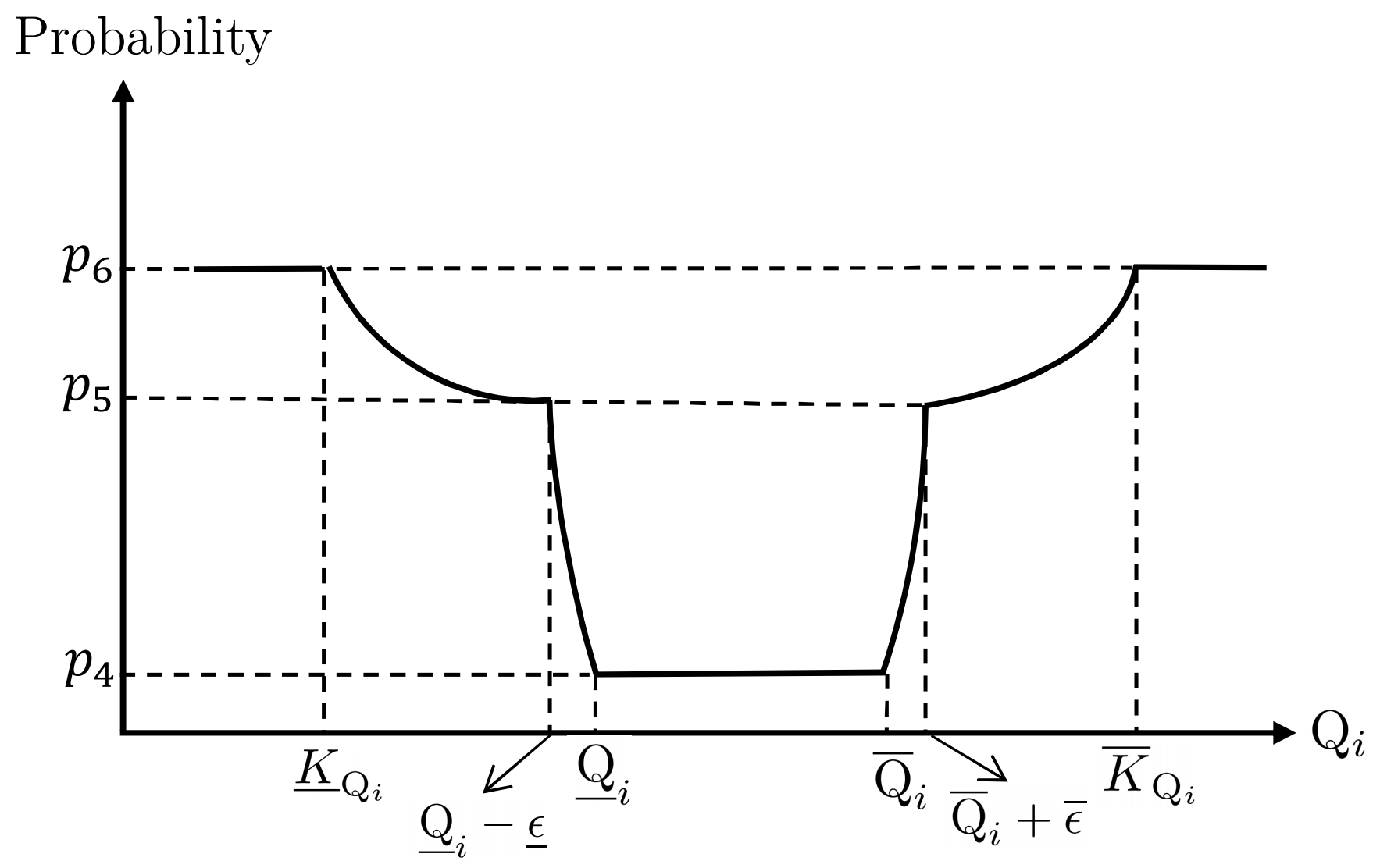}}
\caption{Probability of generator tripping as a function of $Q_i$.}
\label{exponentialg}
\end{figure}

\section{Modeling Protection and Control Strategies} \label{control_sec}

\subsection{Undervoltage Load Shedding} \label{undervoltage} 

When the voltage of a load bus $i$ is below a threshold $\mathrm{V}_\mathrm{th}$ for more than $\tau$ seconds, a portion of the real power load will be shed \cite{otomega2007undervoltage}. 
The amount of active power to be shed is determined as: 
\begin{align}\label{load_s}
\Delta \mathrm{P}^{\mathrm{sh}}_i=\min(K_\mathrm{sh}\Delta V_i,\mathrm{P}_i),
\end{align}
where $\Delta V_i=V_{\mathrm{th}}-V_i>0$.
In this paper, the parameters are chosen as $V_\mathrm{th}=0.9$ pu, $K_{\mathrm{sh}}=600$ MW/pu, 
and $\tau=3$ seconds. 
In order to preserve the power factor, the reactive power to be shed is calculated as \cite{van2015test}:
\begin{align}\label{load_q}
\Delta \mathrm{Q}^{\mathrm{sh}}_i=\mathrm{Q}_i \frac{\Delta \mathrm{P}_i^{\mathrm{sh}}}{\mathrm{P}_i}.
\end{align}

\subsection{Operator Re-Dispatch}\label{control_strategy}

Note that the operator only has the initial branch flow capacity $\overline{F}^0_{ij}$. 
If the branch flow is greater than the dynamic rating but smaller than $\overline{F}^0_{ij}$, the operator will not perform any re-dispatch. Only when the line flow is higher than $\overline{F}^0_{ij}$, will the operator be able to perform a shift-factor ($S$) based re-dispatch, which well reflects the actual operator behavior. 

Assume branch $i\rightarrow j$ is overloaded, i.e. $F_{ij}>\overline{F}^0_{ij}$. We select $n^+$ generators $\{\mathrm{G}_{P_1},\mathrm{G}_{P_2},\cdots,\mathrm{G}_{P_{n^+}}\}$ with positive shift factors and $n^-$ generators $\{\mathrm{G}_{N_1},\mathrm{G}_{N_2},\cdots,\mathrm{G}_{N_{n^-}}\}$ with negative shift factors. 
For the generators with positive shift factors, without loss of generality, we assume 
\begin{align}
S_{P_1} \geq S_{P_2}\geq \cdots \geq S_{P_{n^+}}.
\end{align}
To reduce the line overloading most effectively, the generators with positive shift factors should be dispatched (i.e., decreasing their outputs) in the order of $\mathrm{G}_{P_1},\mathrm{G}_{P_2},\cdots,\mathrm{G}_{P_{n^+}}$. 
Specifically, for generator $\mathrm{G}_{P_i}$, $i=1,\cdots,n^+$, its real power output is re-dispatched as: 
\begin{align} \label{eq:redispatch}
\mathrm{P}_{P_i}= \mathrm{P}^0_{P_i} + \frac{\eta (\overline{F}^0_{ij} - F_{ij})}{S_{P_i}}.
\end{align}

If the generators with positive shift factors cannot eliminate the overloading, the generators with negative shift factors will be re-dispatched. 
Without loss of generality, we assume
\begin{align}
S_{N_1}\leq S_{N_2}\leq \cdots \leq S_{N_{n^-}}.
\end{align}
These generators are dispatched (i.e., increasing their outputs) in the order of  $\mathrm{G}_{N_1},\mathrm{G}_{N_2},\cdots,\mathrm{G}_{N_{n^-}}$ by a similar approach to that for the generators with positive shift factors. 

If there are multiple overloaded branches, the above re-dispatch will be applied to each of them according to $F_{ij}/F_{ij}^0$. The larger $F_{ij}/F_{ij}^0$ of a branch is, the earlier the overloading of this branch will be dealt with by the re-dispatch of generators. If necessary, multiple rounds of re-dispatch will be executed until the overloading of all branches is eliminated or the number of rounds reaches a limit.

\section{Timing of Events}\label{event_time}

Define a set of events for the $(k+1)$th iteration as $E=\{e_1,e_2,\cdots,e_m\}$ where $m$ is the number of potential events that could happen in iteration $k+1$. The events could be low voltage of a load bus, tripping of a line whose $F_{ij}$ could be smaller or greater than its dynamic rating $\overline{F}^\mathrm{d}_{ij}$, tripping of a generator whose reactive power output is within/outside its lower and upper limits. Each type of events will fail after a specific amount of time which will be decided as follows. 

As mentioned in Section \ref{undervoltage}, when the voltage of a load bus $i$ is below a pre-defined threshold for more than $\tau=3$ s, we shed $\Delta \mathrm{P}^{\mathrm{sh}}_i$ and $\Delta \mathrm{Q}^{\mathrm{sh}}_i$ of load at bus $i$.  The re-dispatch in Section \ref{control_strategy} is assumed  to be be finished in one minute. 

If a line whose $F_{ij}$ is smaller than its dynamic rating $\overline{F}^\mathrm{d}_{ij}$ is tripped,  there is no slow process involved and the line is disconnected by the protective relay after a very short time, which may include the relay operating time and the breaker operating time. We consider this time as $0.2$ s \cite{analysis}, \cite{breaker_l}. 

When $F_{ij}\ge \overline{F}^\mathrm{d}_{ij}$, the line may be tripped for different reasons, such as tree contact caused by a slow process \cite{us2004final,slow}, overheating, or mis-operation of zone 2 and zone 3 distance relays \cite{us2004final}. 
The time of line tripping under different mechanisms can vary significantly. 
For example, in the 2003 U.S.-Canadian blackout the Stuart-Atlanta 345-kV line tripping took 31 minutes due to tree contact. The backup zone 2 and zone 3 relay, however, can operate in a few seconds. The probability of a line with $F_{ij}\ge \overline{F}^\mathrm{d}_{ij}$ to fail can be determined based on Section \ref{line_prob_sec}. 
If such a line $l:i \rightarrow j$ is to fail, 
it fails when its total accumulated overload exceeds a limit $\bar{o}_{ij}$ which represents the condition required for line tripping due to a number of processes such as the overheating of a transmission line or the sagging of the line to vegetation \cite{slow,chemistry}. 
Let $\Delta o_{ij,0}=0$. The time for line $l:i \rightarrow j$ whose $F_{ij}$ is greater than $\overline{F}^\mathrm{d}_{ij}$ in iteration $k+1$ 
can be calculated as:
 \begin{align}\label{tb}
\Delta t_{ij,k+1} = \frac{\bar{o}_{ij}-\sum_{m=0}^{k} \Delta o_{ij,m}}{F_{ij}(t_{k})-\overline{F}^\mathrm{d}_{ij}(t_{k})},
\end{align}
where the limit $\bar{o}_{ij}$ is chosen so that a branch will trip after 20 seconds of being 50\% above the branch flow limit. 

According to \cite{kunder}, generators usually have about 10 to 20\% overload capability for up to 30 minutes. We use a similar approach to determine the time for generator tripping in every interval. When  the  reactive  power  of  a  generator $g$ is in $[\underline{\mathrm{Q}}_g,\overline{\mathrm{Q}}_g]$, it fails by a very small probability $p_1$ because of  accidental failure. 
For these types of failures, we consider the time as $0.2$ second \cite{classes}. If the reactive power of generator $g$ moves out of $[\underline{\mathrm{Q}}_g,\overline{\mathrm{Q}}_g]$, the probability for that generator to fail can also be determined by Section \ref{probab_gen_tripping}. If the generator is to fail, the time that is required for this generator to fail in iteration $k+1$ can be calculated as
\begin{align}
\Delta t_{g,k+1}&= \begin{cases}
\frac{\bar{o}_{g}-\sum_{m=0}^{k}\Delta o_{g,m}}{\underline{\mathrm{Q}}_g-\mathrm{Q}_g(t_{k})},  &\text{if $\,\mathrm{Q}_g < \underline{ \mathrm{Q}}_g$} \\ \\
\frac{\bar{o}_{g}-\sum_{m=0}^{k} \Delta o_{g,m}}{{\mathrm{Q}_g(t_{k})-\overline{\mathrm{Q}}}_g}, &\text{if $\,\mathrm{Q}_g > \overline{\mathrm{Q}}_g$}, \\
\end{cases}
\end{align}
where the threshold $\bar{o}_g$ is chosen so that a generator will trip after 30 minutes of being 20\% above/below the upper/lower reactive power limit and $o_{g,0}=0$.

Let $\Delta t_{k+1}^{\min}$ be the minimum time of all events in iteration $k+1$, and can be calculated as
\begin{align}
\Delta t_{k+1}^{\min}& = \min\Big\{t(e_i),i=1,\cdots,m \Big\}, 
\end{align}
where $t(e_i)$ is the time for event $e_i$ and the time corresponding to the next event is thus $t_{k+1}=t_k+\Delta t_{k+1}^{\min}$. 
 
The $\Delta o_{ij}$ at iteration $k+1$ can be obtained by \cite{chemistry}:
 \begin{align}
\Delta o_{ij,k+1}&= \max\Big(F_{ij}(t_k)-\overline{F}^\mathrm{d}_{ij}(t_{k}),0\Big)\Delta t_{k+1}^{\min},
\end{align}
and $\Delta o_{g}$ at iteration $k+1$ can be calculated as 
\begin{align}
\Delta o_{g,k+1}&= \begin{cases}
\Big(\underline{\mathrm{Q}}_g-\mathrm{Q}_g(t_{k})\Big)\Delta t_{k+1}^{\min},  &\text{if $\, \mathrm{Q}_g < \underline{\mathrm{Q}}_g$}\\\\
\Big(\mathrm{Q}_g(t_k)-\overline{\mathrm{Q}}_g\Big)\Delta t_{k+1}^{\min},  &\text{if $\, \mathrm{Q}_g > \overline{\mathrm{Q}}_g$}. \notag 
 \end{cases}
 \end{align}

\section{Voltage Stability Margin Calculation} \label{vsm_intro}

Voltage instability has been responsible for several major blackouts, such as New York Power Pool disturbance on September 22, 1970 and Western systems coordination council (WSCC) transmission system disturbance on July 2, 1996. 

A system enters a state of voltage instability when a disturbance, such as a load increase or change in system conditions, causes a progressive and uncontrollable decline in voltage. 
 In blackouts, the load increase due to temperature disturbance and the reactive power supply decrease due to the increased probability of tripping of the lines and generators that are geographically close to the load increase area both contribute to voltage instability. 
More importantly, the load increase and the increase of line tripping probability are correlated and  both are related to the temperature disturbance, which may greatly increase the risk of cascading. 

After each change in the operating condition, we calculate the voltage stability margin based on the QV index proposed in \cite{index}. 
Specifically, for a power flow model
\begin{equation} \label{pf}
\renewcommand{\arraystretch}{1.0}
\left[\begin{array}{c}
\Delta \mathrm{P}  \\
\Delta \mathrm{Q}
\end{array}\right] = \left[ \begin{array}{cc}
J_{\mathrm{P}\theta} & J_{\mathrm{P}\mathrm{V}} \\
J_{\mathrm{Q}\theta} & J_{\mathrm{Q}\mathrm{V}}
\end{array} \right]
\left[\begin{array}{c}
\Delta \theta \\
\Delta \mathrm{V}
\end{array}\right],
\end{equation}
letting $\Delta \mathrm{P}=0$ we have
\begin{align} \label{theta}
\Delta \theta = - J_{\mathrm{P}\theta}^{-1} J_{\mathrm{PV}} \Delta \mathrm{V}.
\end{align}
Substituting (\ref{theta}) into the $\Delta \mathrm{Q}$ equations in (\ref{pf}) we have
\begin{align}
\Delta \mathrm{Q} = \left( J_{\mathrm{QV}} - J_{\mathrm{Q}\theta} J^{-1}_{\mathrm{P}\theta} J_{\mathrm{PV}} \right) \Delta \mathrm{V}.
\end{align}
Let $J_R=J_{\mathrm{QV}} - J_{\mathrm{Q}\theta} J^{-1}_{\mathrm{P}\theta} J_{\mathrm{PV}}$ and a voltage stability index (VSI) for the whole system can be defined as
\begin{align}\label{VSIIndex}
\mathrm{VSI}=\min \Bigg\{\frac{\det (J_R)}{\big[\mathrm{adj}(J_R)\big]_{ii}}, i=1,...,N \Bigg\}, 
\end{align}
where $N$ is the nubmer of buses, $\mathrm{adj}(A)=\det(A)A^{-1}$, and $\det(A)$ is the determinant of $A$.
The VSI can be used to indicate how close the system is to voltage instability. The bigger  VSI is, the more stable the system is. When VSI approaches zero the system will lose voltage stability \cite{index}.

\section{Proposed Blackout Model} \label{model-summary}

\begin{figure}[!t]
\centering
\centerline{\includegraphics[width=0.95\columnwidth]{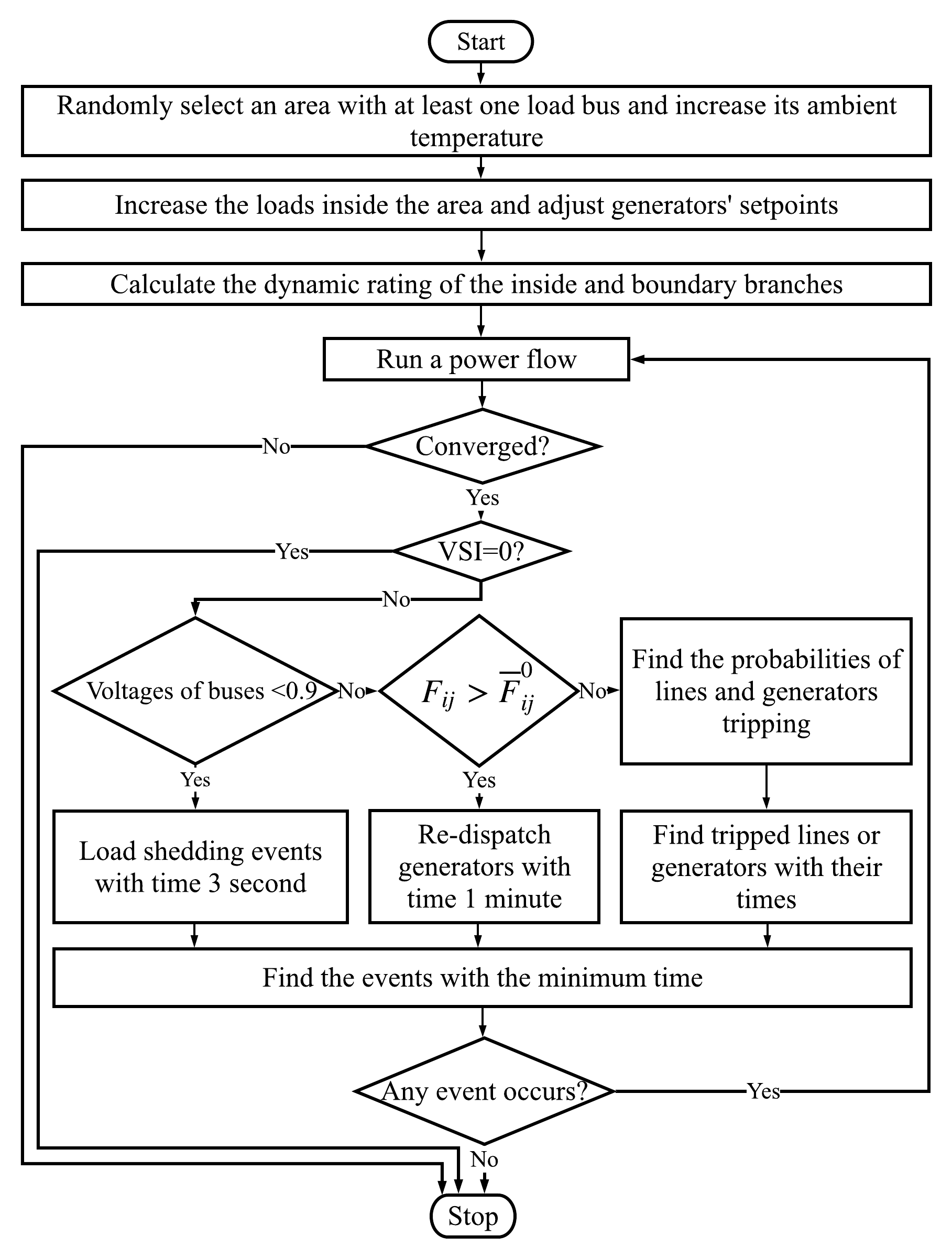}}
\captionsetup{justification=raggedright,singlelinecheck=false}
\caption{Flowchart of the proposed blackout model. }
\captionsetup{justification=raggedright,singlelinecheck=false}
\label{flowchart}
\end{figure}

Fig. \ref{flowchart} illustrates the proposed cascading failure model which can be implemented in the following ten steps.

\begin{enumerate}
	\item Randomly select a load bus and an area around the selected load bus based on Section \ref{tem_dist}. Increase the ambient temperature of the selected area by $\Delta T$.
	\item Increase the loads of the buses inside the selected area based on Section \ref{load_temp_subsec} and adjust the generators' real power setpoints based on (\ref{P_change}) in Section  \ref{probab_gen_tripping}.
	\item Calculate dynamic ratings of the lines inside the selected area and the boundary lines by \eqref{dynamic_rating_formula1} in Section \ref{dynamic_rating_sec}.
	\item \label{run_pf} Run power flow. If the power flow cannot converge, go to Step \ref{final}; if the power flow converges but there is no voltage stability margin as defined in \eqref{VSIIndex}, go to Step \ref{final}. Otherwise, go to Step \ref{shed}. 
	\item\label{shed} Check the voltage of the load buses and the load for the load buses whose voltages are less than $V_\mathrm{th}$ is to be shed based on Section \ref{undervoltage}. 
	\item\label{redispatch} Check line flows. If there is any line whose line flow $F_{ij}$ is greater than its initial rating $F_{ij}^0$, the generators are to be re-dispatched based on Section \ref{control_strategy}. 
	\item\label{probabl}Find the probabilities of line tripping from Section \ref{line_prob_sec} and generator tripping from Section \ref{probab_gen_tripping} and decide the lines and generators that will be tripped.
	\item \label{events} Calculate the time for each event according to Section \ref{event_time}. The event from Steps \ref{shed}--\ref{probabl} that has the minimum time will actually happen. 
	\item If any event occurs, go back to Step \ref{run_pf}; otherwise, go to Step \ref{final}. 
	\item \label{final} Stop the simulation. 
\end{enumerate}

By utilizing our proposed model, we can realistically model cascading failures and capture what has happened in previous blackouts. Even though the system is $N-1$ secure, cascading failure can still be initiated and then propagates in a large area of the system. This is mainly due to the following reasons. 
\begin{enumerate}
    \item Although load forecasting is used for day-ahead generation purchases and reactive power management,  due to uncertainties such as unexpected temperature disturbances the actual load may be different from the forecasted load, thus changing system operating conditions. 
    \item Under different weather conditions the actual line rating could change significantly due to a number of processes such as the overheating of a transmission line or the sagging of the line to vegetation. An $N-1$ secure system under initial line ratings may not still be $N-1$ secure under the reduced dynamic line ratings such as due to temperature increases. 
    \item The initiating events have important geographical correlations due to external weather conditions such as temperature disturbance. Temperature increase can cause increased line flow by load increase and line rating decrease at the same time, greatly increasing the probability of tripping of lines inside or on the boundary of an area with temperature disturbances. Generators inside the area with temperature disturbance also have increased chance of being disconnected due to load increase and also reduction of reactive supply from the outside system after tie line disconnection. Our cascading failure model can take into account the correlations between different events such as line outage, generator tripping, and undervoltage of load buses, which may lead to extensive outage propagation even if the system is initially $N-1$ secure without any temperature disturbance.
\end{enumerate}

\section{Results} \label{result}

The proposed model is implemented in Matlab for the RTS-96 3-area system \cite{grigg1999ieee} based on MATPOWER \cite{matpower}. There are 73 buses and 120 branches in the system, and the total load is 8550 MW. Compared to the initial model, each reactor of 100 MVAr at buses 106, 206, and 306 is split into two, 50 MVAr at each extremity of the cable. These reactors are considered to be automatically disconnected in case of the outage of the corresponding cable. The pre-contingency steady state is based on a Preventive-Security-Constrained Optimal Power Flow so that the system is $N-1$  secure\footnote{\url{http://homepages.ulb.ac.be/~phenneau/CFWG\_Benchmark.HTML}}\cite{henneaux2018benchmarking}. 
All tests are carried out on a 3.20 GHz Intel(R) Core(TM) i7-8700 based desktop.

\subsection{Parameter Settings}

We set $k^{\mathrm{pf}} = 0.001$ for all buses \cite{us2004final}, $K=1.5$ in Section \ref{line_prob_sec} for lines\cite{lee2008estimating}, $\underline{K}_{\mathrm{Q}_i}$ in Section \ref{probab_gen_tripping} as $1.5\underline{\mathrm{Q}}_i$ if $\underline{\mathrm{Q}}_i<0$ and as $-0.5$ if $\underline{\mathrm{Q}}_i=0$, and $\overline{K}_{\mathrm{Q}_i}$ as $1.5\overline{\mathrm{Q}}_i$. 
Besides, $\epsilon$ is chosen as $0.01$ for lines \cite{lee2008estimating}, and $\underline{\epsilon}=-0.01 \underline{\mathrm{Q}}_i$ and $\overline{\epsilon}=0.01\overline{\mathrm{Q}}_i$ for generators.  
We choose $p_1=0.001$,  $p_2=0.3$, and $p_3=1$ \cite{lee2008estimating}. For generators we choose $p_4=0.001$ to consider hidden failures, $p_5=0.3$, and $p_6=1$. For re-dispatch we set $\eta=1.05$ \cite{yao2016}. For calculating dynamic line rating, we consider $\alpha_{ij}=1$.

\subsection{Typical Simulation Run without Operator Re-Dispatch} \label{typical_case}

Buses 207 and 208 are selected as the internal buses and their initial temperatures are set to be $T_{\mathrm{high}}^0=24.21^{\circ}\mathrm{C}$. As a temperature disturbance, we increase the temperature of the selected area to  $T=T_{\mathrm{high}}^0+10^{\circ}\mathrm{C}$. As in Table \ref{table:initial}, after temperature increase the active and reactive powers at these two buses also increase. The line flows and the dynamic line ratings of the boundary and internal branches are listed in Table \ref{table:flows}. After temperature rise the line flows increase while $\overline{F}_{ij}^\mathrm{d}$ decreases compared to $ \overline{F}^0_{ij}$, and branches 207--208 and 208--209 become overloaded.

\begin{table}[htb!]
\centering
\footnotesize
\captionsetup{labelsep=space,font={footnotesize,sc}}
\caption{\\ Internal Buses with Their Active and Reactive Powers in the Typical Case without Operator Re-Dispatch}
\begin{tabular}{c c c c c}
\hline\hline \xrowht{8pt}
\begin{tabular}[c]{@{}c@{}}Inside\\  buses\end{tabular} & \begin{tabular}[c]{@{}c@{}}$\mathrm{P}(T_i^0)$\\ (MW)\end{tabular} & \begin{tabular}[c]{@{}c@{}}$\mathrm{P}(T_i)$\\ (MW)\end{tabular} & \begin{tabular}[c]{@{}c@{}}$\mathrm{Q}(T_i^0)$\\ (MVAr)\end{tabular} & \begin{tabular}[c]{@{}c@{}}$\mathrm{Q}(T_i)$\\ (MVAr)\end{tabular}\\ \hline 
 207 & 125  &  189.79  &  25  & 48.08  \\ 
  208 & 171  &  259.63  &  35  & 66.76   \\
 \hline\hline
\end{tabular}
\label{table:initial}
\end{table}

\begin{table}[htb!]
\centering
\scriptsize
\captionsetup{labelsep=space,font={footnotesize,sc}}
\caption{\\Initial Flows, Flow after Temperature Rises, and Dynamic L\sc{ine} Rating of Boundary and Internal Branches in the Typical Case without Operator Re-Dispatch} 
\label{flows}
 \resizebox{0.4\textwidth}{!}{%
\begin{tabular}{c c c c c c}
\hline\hline \xrowht{8pt}
Branch  &$F_{ij}(T_{ij}^0)$ &  $\overline{F}^0_{ij}$ &$F_{ij}(T_{ij})$ 
 & $\overline{F}_{ij}^\mathrm{d}(T_{ij})$ \\ \hline \xrowht{6pt}
(207, 208)  & 53.24  &175  &   148.89 & 147.40 
 \\ \xrowht{6pt}
 \bf{(208, 209)}  & \bf{96.41} & \bf{190}  & \bf{175.07} &   \bf{173.21} 
 \\ \xrowht{6pt}
 (208, 210)  & 82.77  & 190  &  161.81 &173.45 
 \\ 
  \hline\hline
\end{tabular}}
\label{table:flows}
\end{table}

\begin{figure}[!t]
	\centering
	\centerline{\includegraphics[width=2.9in]{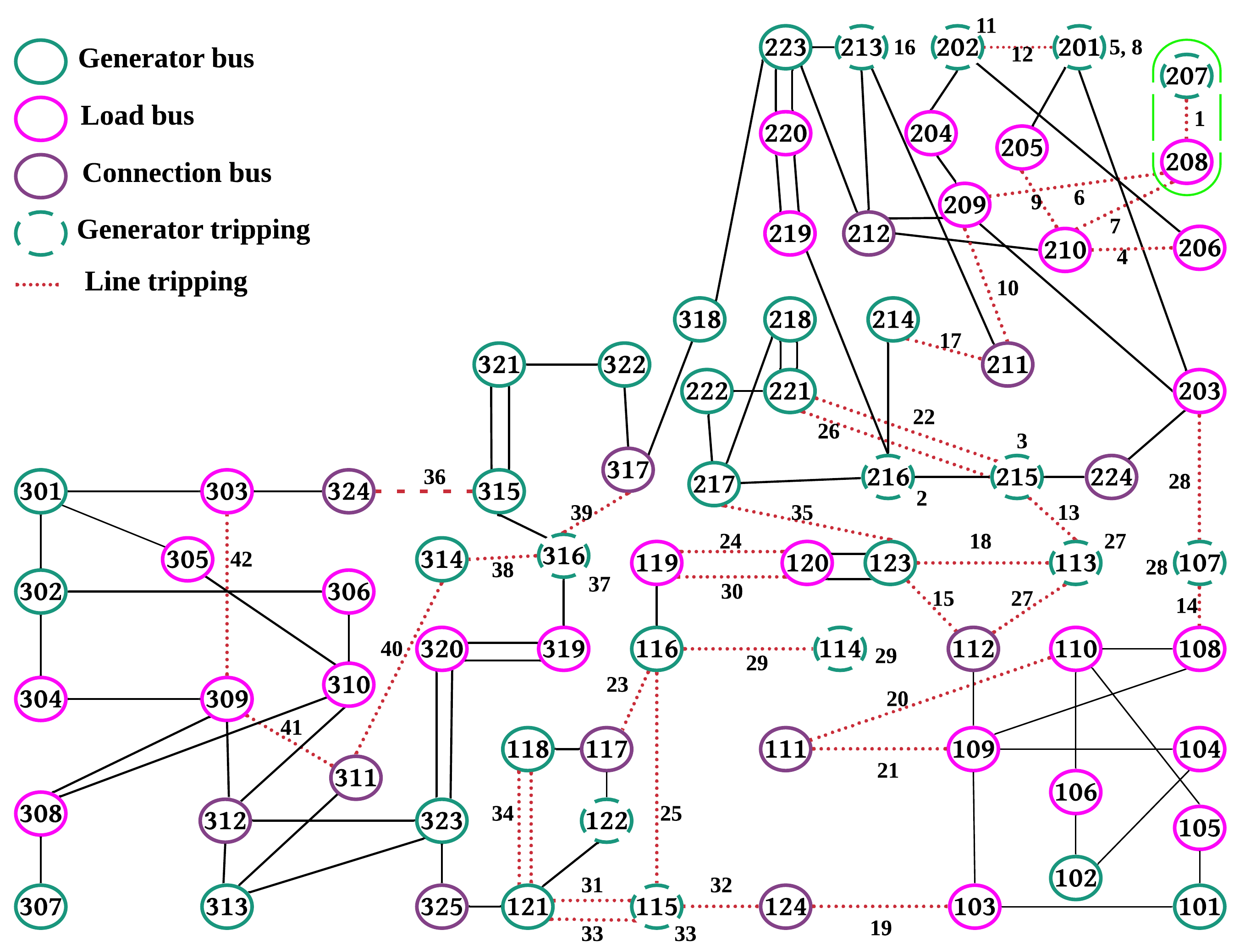}}
	\captionsetup{justification=raggedright,singlelinecheck=false}
	\caption{Process of a typical blackout in RTS-96 system simulated by the proposed model without operator re-dispatch.}
	\label{event}
\end{figure}

\begin{figure}[!t]
	\centering
	\centerline{\includegraphics[width=3.0in]{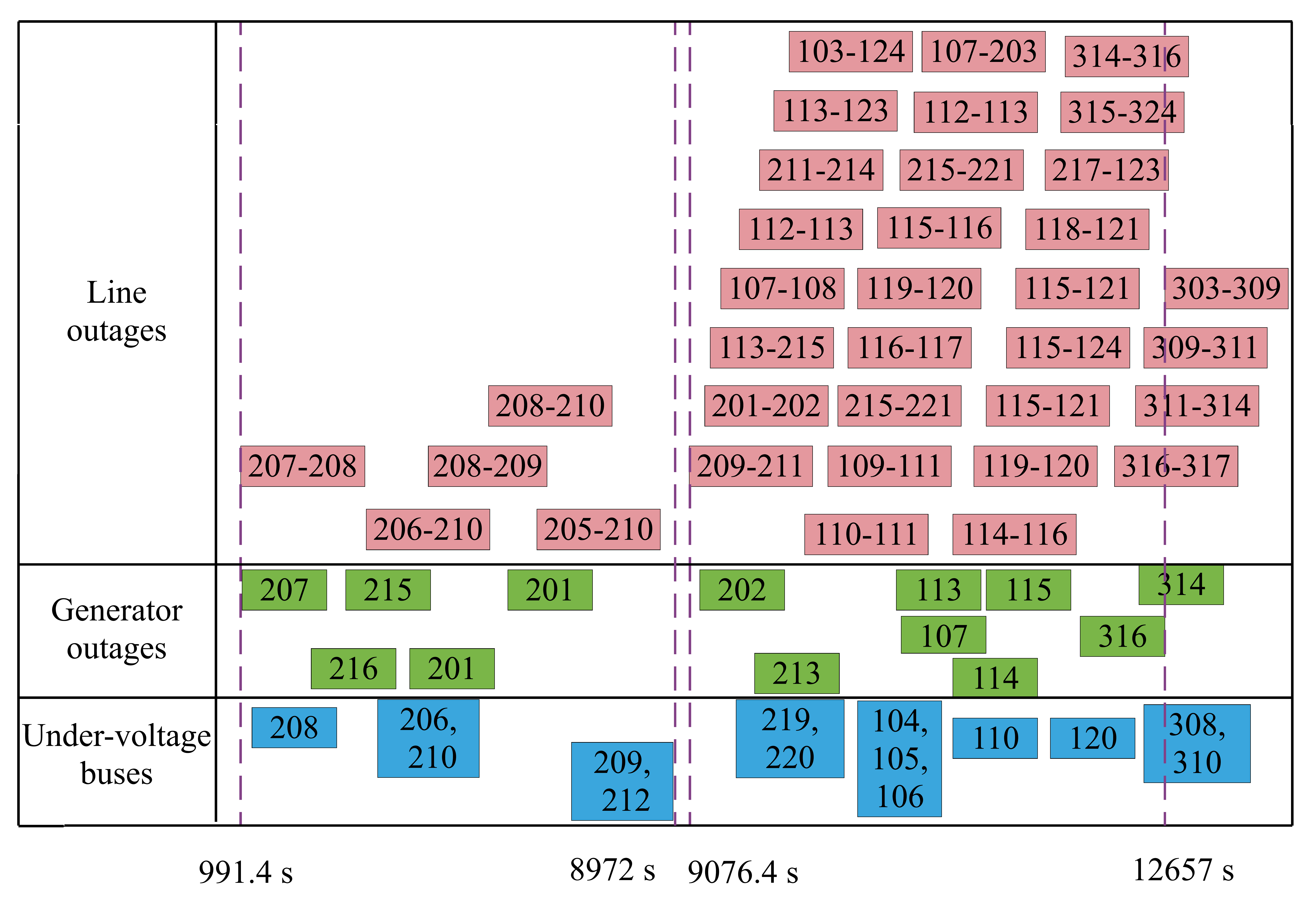}}
	\captionsetup{justification=raggedright,singlelinecheck=false}
	\caption{The events for the typical case without operator re-dispatch. For generator outages the involved generator buses are shown. Since one generator bus can have several generators connected to it, it may appear more than once, such as generator bus 201.}
	\label{event_black}
\end{figure}

If dynamic line rating is not considered, ${F}_{ij}(T_{ij})$ for all lines are less than $ \overline{F}^0_{ij}$ and the probability for line tripping is as low as $p_1$ and the probability for generator tripping is equal to $p_4$. However, if dynamic line rating is considered, things will be totally different.  
When operator re-dispatch is not modeled, the event sequence simulated from the proposed model is shown in Fig. \ref{event}, in which line tripping is indicated by red dash lines, generator tripping is indicated by green dash circles, and the number next to the tripped line or generator indicate the sequence of the event. After the temperature of the selected area increases, branch 207--208 will be tripped by probability $0.3$. In our simulation it is tripped after $991.40\,\mathrm{s}$. This leads to undervoltage of bus $208$ and the islanding of the generators at bus 207. Then the cascading  gradually propagates to the other parts of the system, leading to a total of 35 line outages and tripping of 19 generator  at 12 generator buses. The line outages, generator outages, and undervoltage buses during the blackout are shown in Fig. \ref{event_black}.

\begin{figure}[!t]
	\centering
	\centerline{\includegraphics[width=2.9in]{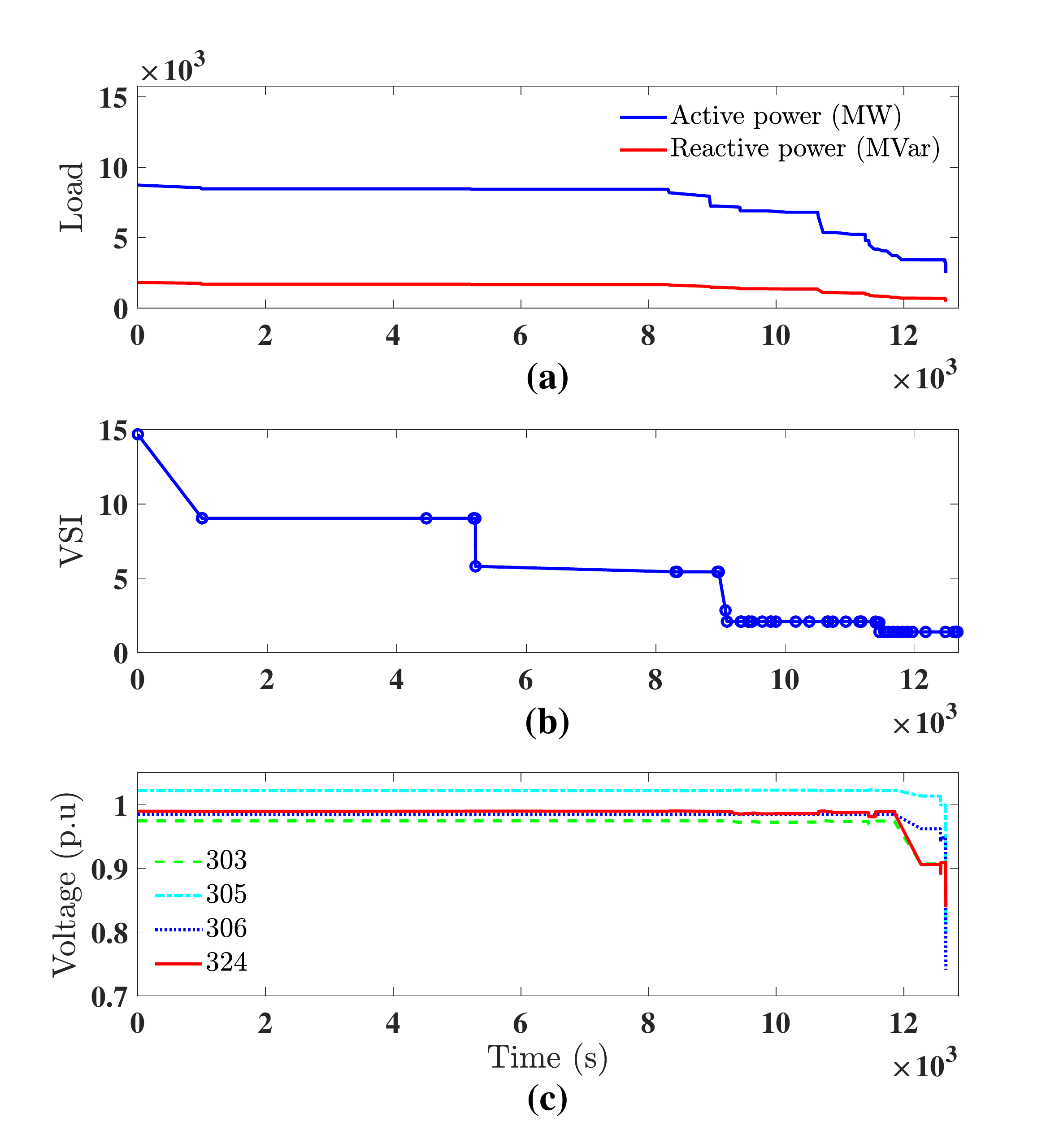}}
	\captionsetup{justification=raggedright,singlelinecheck=false}
	  \caption{Total load, VSI, and voltage at vulnerable buses for the typical case without operator re-dispatch: (a) Total load; (b) VSI; (c) Voltage at vulnerable buses.} 
	\label{threefigure}
\end{figure}

The total load during the blackout is shown in Fig. \ref{threefigure}(a). Three seconds after the first line tripping, part of the load at bus 208 is shed due to undervoltage. Due to islanding and undervoltage, load shedding becomes much faster after 8036 seconds. The VSI during the cascading event is shown in Fig. \ref{threefigure}(b). Under normal operating conditions VSI is 14.68, and when the blackout propagates it gradually decreases. At 11454.2 s, VSI decreases from 2.02 to 1.39. Fig. \ref{threefigure}(c) shows the voltages of four vulnerable buses (buses 303, 305, 306, and 324) over time. It is seen that after the VSI reduces to  1.39 the voltages at the vulnerable buses begin to drop significantly before voltage collapse finally occurs at 12657 s.

\subsection{Typical Simulation Run with Operator Re-Dispatch}

\begin{figure}[!t]
	\centering
	\centerline{\includegraphics[width=3.0in]{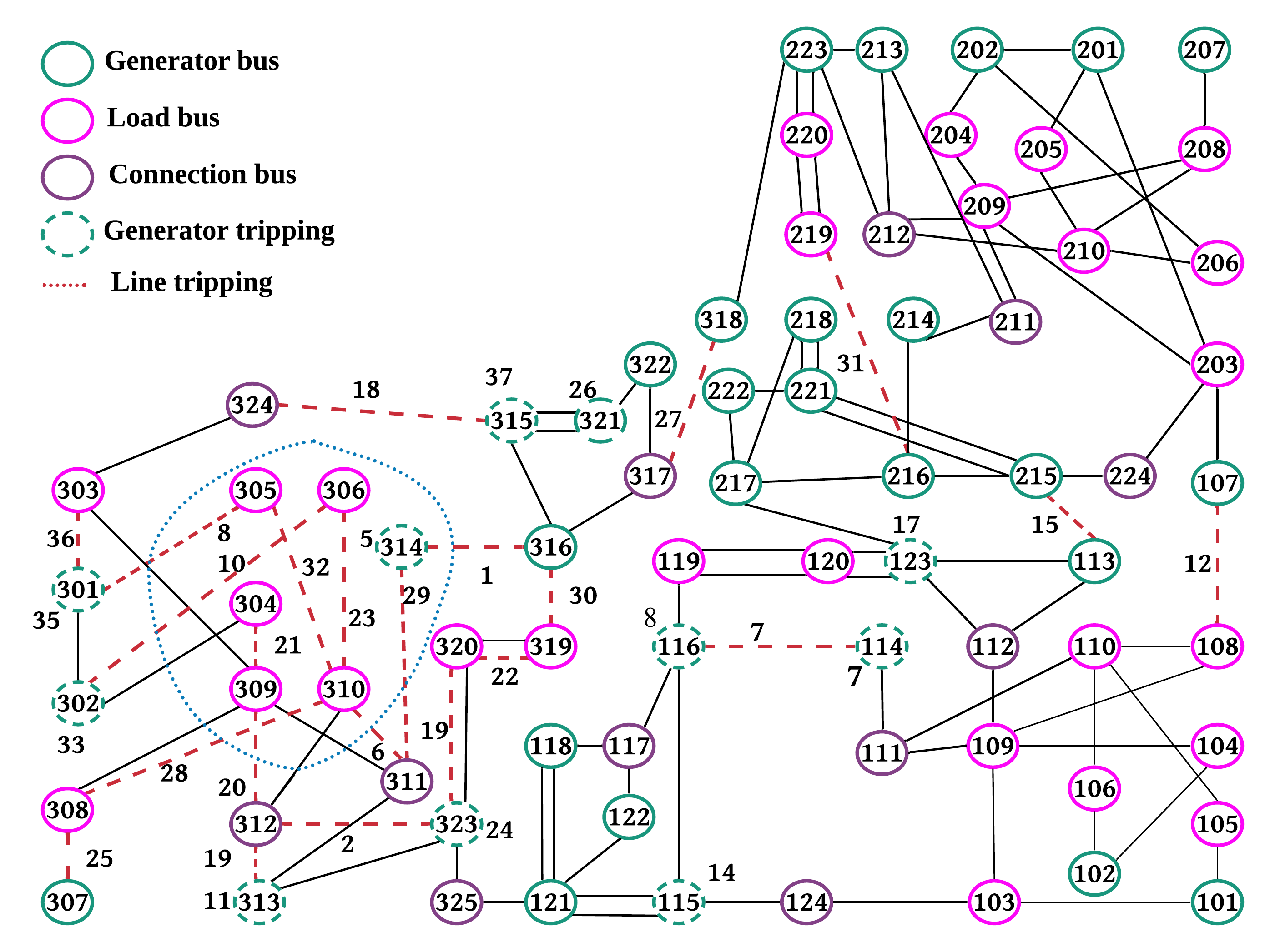}}
	\captionsetup{justification=raggedright,singlelinecheck=false}
	\caption{Process of a typical blackout in RTS-96 system when operator re-dispatch is modeled.}
	\label{event2}
\end{figure}

\begin{figure}[!t]
	\centering
	\centerline{\includegraphics[width=2.9in]{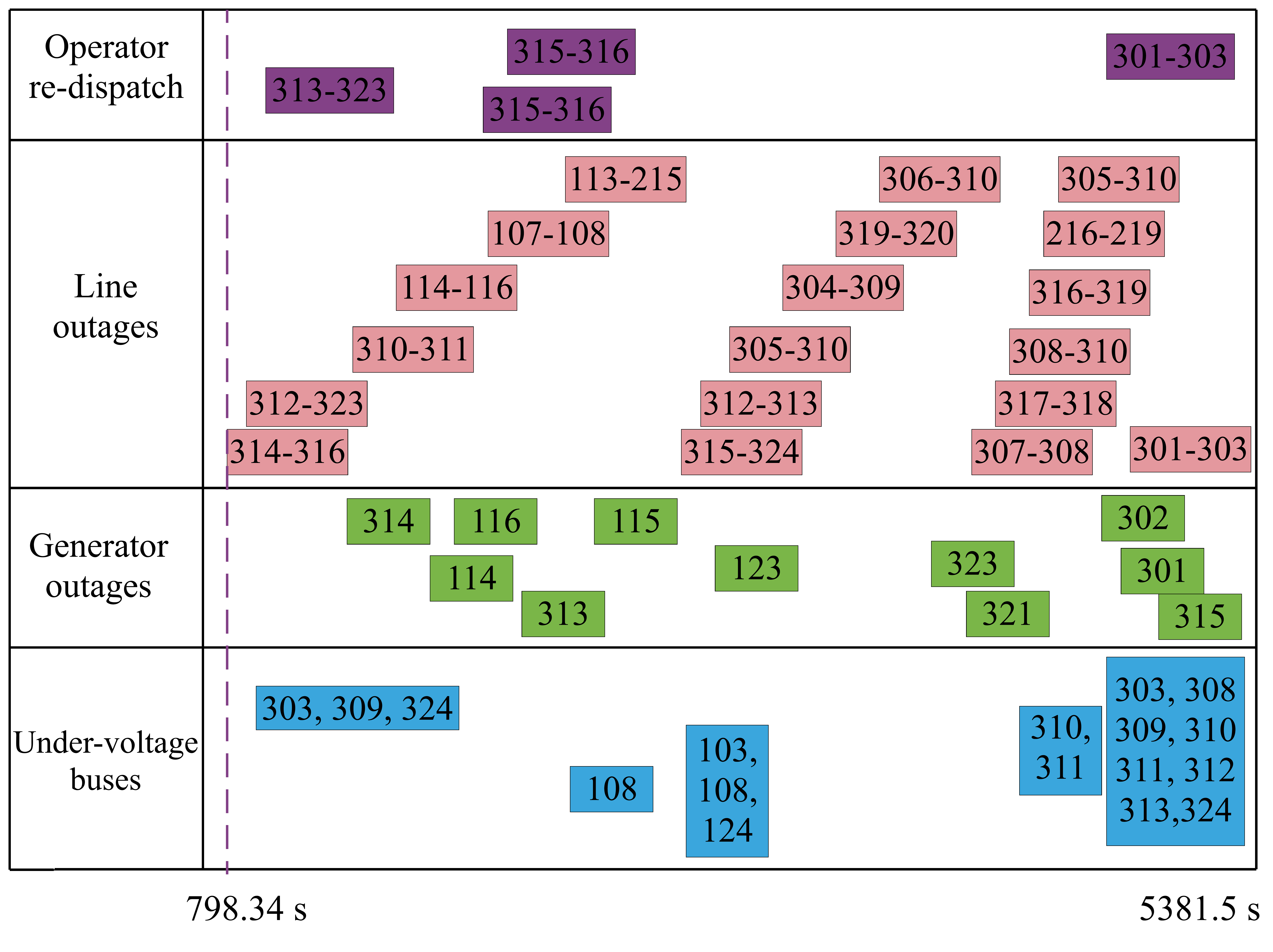}}
	\captionsetup{justification=raggedright,singlelinecheck=false}
	\caption{The events for the typical case with operator re-dispatch. For generator outages the involved generator buses are shown.}
	\label{event_black2}
\end{figure}

In this case, buses 304, 305, 306, 309, 310, and 314  are chosen as the internal buses and their initial temperatures are $T_{\mathrm{high}}^0=24.21^{\circ}\mathrm{C}$. The temperature of the selected area is raised to $T=T_{\mathrm{high}}^0+10^{\circ}\mathrm{C}$ to simulate a temperature disturbance. As shown in Table \ref{table:initial1}, the active and reactive powers at the internal buses increase due to temperature rise. The line flows and dynamic line ratings of the boundary and internal branches are reported in Table \ref{flowstypical2}. It is easily seen that the line flow of branch 314--316 becomes higher than its dynamic line rating and is to be tripped by probability $0.32$. In our simulation, it is tripped after 798.34 s and initiates a cascading failure. 
The event sequence simulated from the proposed model with operator re-dispatch is illustrated in Fig. \ref{event2}, in which the selected area is denoted by blue dash lines, line tripping is denoted by red dash lines, generator tripping is specified by green dash circles, and the number next to the tripped line or generator indicates the sequence of the event. The tripping of line 314--316 leads to the overloading of line 312--323 which is tripped after 54.2 s. These events result in the undervolatge of buses $303$,  $309$, and $324$ at 855.54 s and the overloading of line 313--323 at 915.54 s. The operator re-dispatch at 975.54 s eliminates the overloading of line 313--323, but the generator at bus 314 within the selected area is tripped at 2415.75 s due to over-excitation and cascading failure spreads to the other parts of the system, causing a total of 19 line outages and the tripping of 14 generators at 11 generator buses. The operator re-dispatch, line outages, generator outages, and undervoltage buses during the blackout are illustrated in Fig. \ref{event_black2}.

\begin{table}[htb!]
\centering
\footnotesize
\captionsetup{labelsep=space,font={footnotesize,sc}}
\caption{\\ Internal Buses with Their Active and Reactive Powers in the Typical Case with Operator Re-Dispatch}
\begin{tabular}{c c c c c}
\hline\hline 
\begin{tabular}[c]{@{}c@{}}Inside\\  buses\end{tabular} & \begin{tabular}[c]{@{}c@{}}$\mathrm{P}(T_i^0)$\\ (MW)\end{tabular} & \begin{tabular}[c]{@{}c@{}}$\mathrm{P}(T_i)$\\ (MW)\end{tabular} & \begin{tabular}[c]{@{}c@{}}$\mathrm{Q}(T_i^0)$\\ (MVAr)\end{tabular} & \begin{tabular}[c]{@{}c@{}}$\mathrm{Q}(T_i)$\\ (MVAr)\end{tabular}\\ \hline 
    304 & 74   &  112.4  &  15  & 29.9  \\ 
  305  & 71   &  107.8  &  14  & 27.4   \\ 
   306 & 141  &  214.1  &  28  & 54.7 \\ 
   309 & 175  &  265.7  &  36  & 69.5  \\ 
   310 & 195  &  296.1  &  40  & 77.3  \\ 
  314  & 194  &  294.6  &  39  & 75.9   \\ 
 \hline\hline
\end{tabular}
\label{table:initial1}
\end{table}

\begin{table}[htb!] 
\centering
\scriptsize
\renewcommand{\arraystretch}{1.2}
\captionsetup{labelsep=space,font={footnotesize,sc}}
\caption{\\Initial Flows, Flow after Temperature Rises, and Dynamic L\sc{ine} Rating of Boundary and Internal Branches in the Typical Case with Operator Re-Dispatch}
 \resizebox{0.42\textwidth}{!}{%
\begin{tabular}{c c c c c c}
\hline\hline \xrowht{9pt}
Branch  &$F_{ij}(T_{ij}^0)$  & $\overline{F}^0_{ij}$ & $F_{ij}(T_{ij})$
 & $\overline{F}_{ij}^\mathrm{d}(T_{ij})$ \\  \hline  \hline   
 (305, 301)   &  48.3  & 175  & 88.1 &   167.7 \\  
  (304, 302)   &  39.4  & 175 & 67.3 &   172.5 \\ 
 (306, 302)   & 41.1 & 175  & 66.5 & 160.2 \\ 
 (309, 303) &  20.8  & 175  &  40.9  &   158.4 \\ 
 (304, 309) &  49.2  &175   &  82.7  &   147.4 \\ 
   (305, 310)  &  25.3 & 175  &  	47.4 &   147.4 \\ 
    (306, 310)   &  150.4 & 180 &  151.6 &   152.4 \\ 
     (309, 308)   &  97.4 & 190 &  81.8 &   174.3\\ 
      (309, 311)   &    146.5 &400  &  191.1 &   352.5\\ 
     (309, 312)  & 165.1 & 400  & 211.2  &   353.7 \\ 
       (310, 311)   &  193.4 & 400 &  266.3 &   351.3\\ 
     (310, 312)  &  216.5  & 400 & 288.2  &   342.5 \\ 
  \bf{(314, 316)}    & \bf{354.7}& \bf{500}   &  \bf{498.5} &   \bf{492.4} \\ 
  \hline\hline
\end{tabular}}
\label{flowstypical2}
\end{table}

\begin{figure}[!t]
	\centering
	\centerline{\includegraphics[width=2.9in]{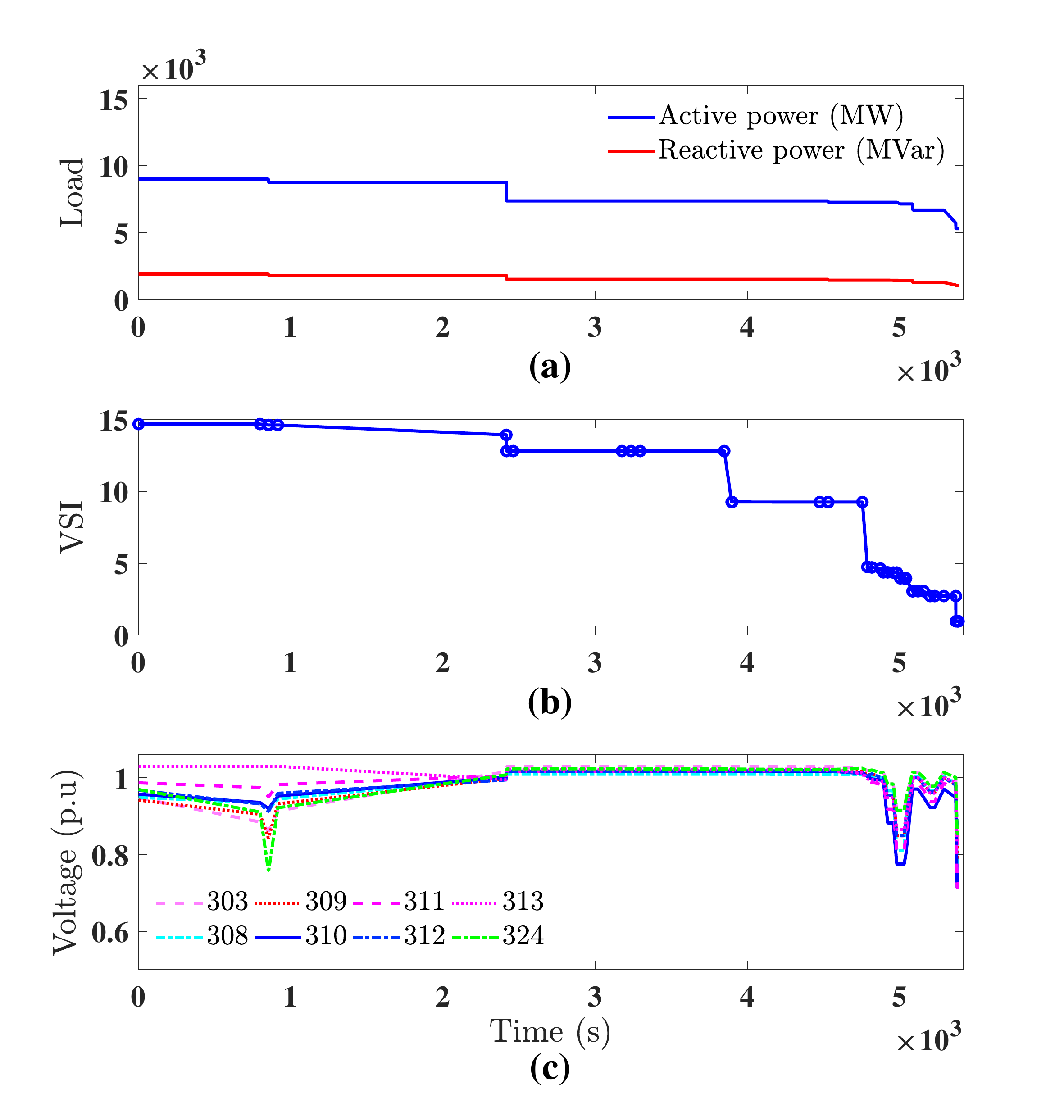}}
	\captionsetup{justification=raggedright,singlelinecheck=false}
	  \caption{Total load, VSI, and voltage at vulnerable buses for the typical case with operator re-dispatch: (a) Total load; (b) VSI; (c) Voltage at vulnerable buses.}
	\label{threefigure2}
\end{figure}

The total load during the blackout is shown in Fig. \ref{threefigure2}(a). After two line trippings, undervoltage happens at buses $303$, $309$, and $324$, and part of the load at these buses is shed due to undervoltage. 
The VSI during the cascading event is shown in Fig. \ref{threefigure2}(b).  
At 5378.5 s, VSI drops from 2.72 to 0.97. Fig. \ref{threefigure2}(c) illustrates the voltages of eight vulnerable buses (buses 303, 308, 309, 310, 311, 312, 313, and 324), some of which are inside the selected area. It is noticed that after VSI declines to 0.97 the voltages at the vulnerable buses begin to drop remarkably before voltage eventually collapses at 5381.5 s.

\subsection{Number of Simulations}

\begin{figure}[!t] 
\centering
\begin{subfigure}{.24\textwidth}
  \centering
  \includegraphics[height=1.3in, width=1.7in]{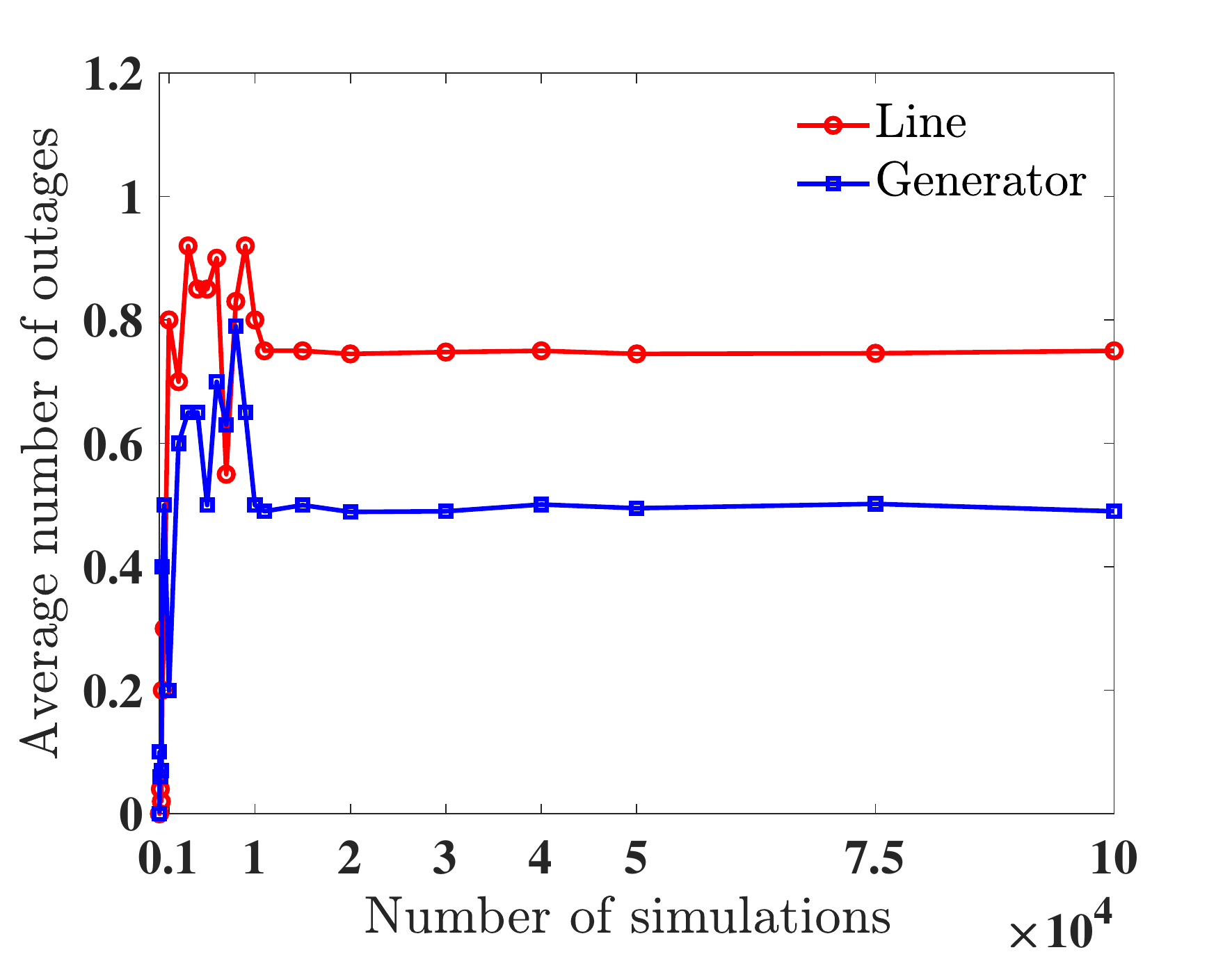}
  \caption{}
\end{subfigure}%
\begin{subfigure}{.24\textwidth}
  \centering
  \includegraphics[height=1.3in, width=1.7in]{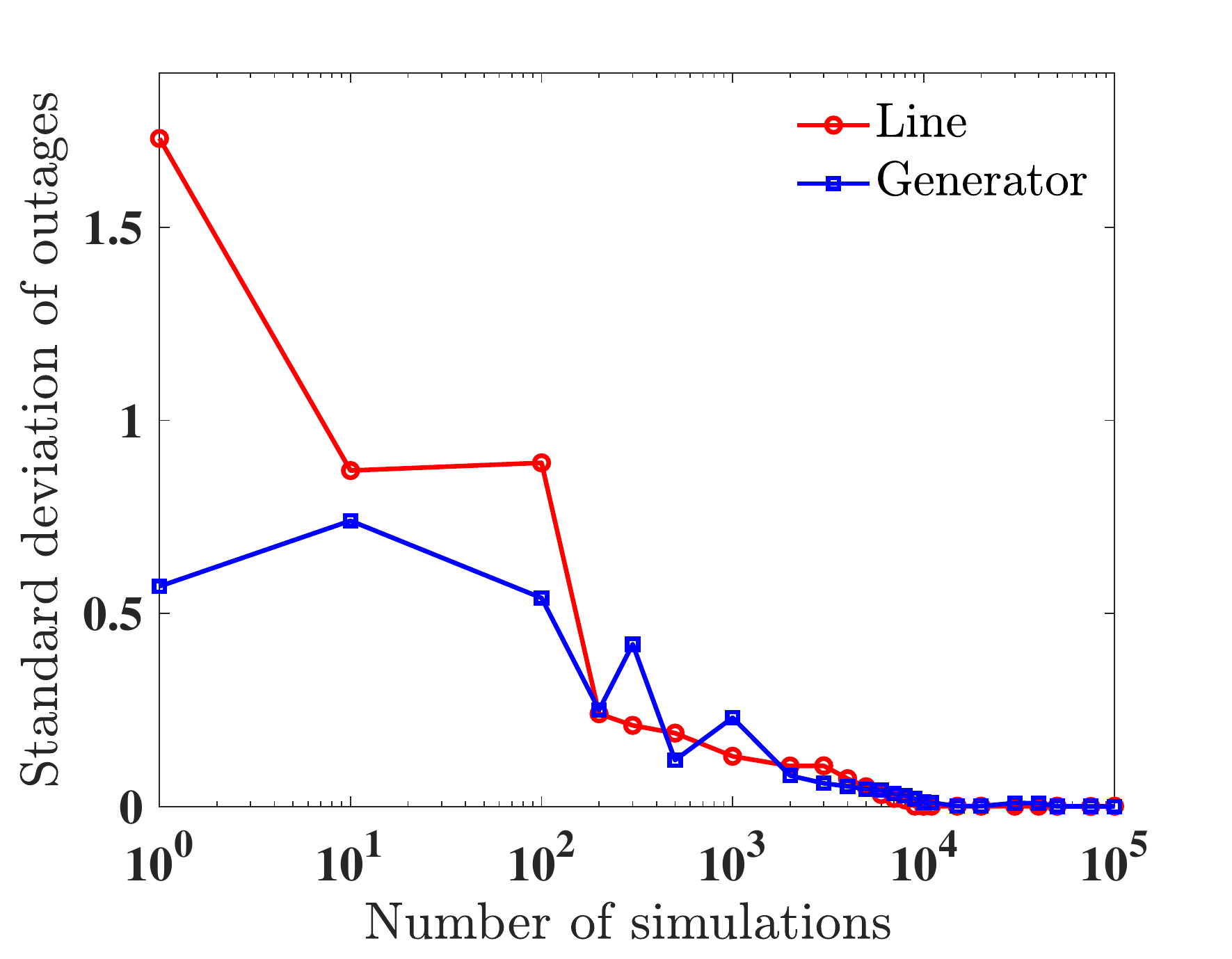}
  \caption{}
  \label{std}
  \end{subfigure}
  	\captionsetup{justification=raggedright,singlelinecheck=false}
 \caption{Average value and standard deviations of the number of outages for different number of simulations: (a) Average value; (b) Standard deviation.}
 \label{num}
  \end{figure}

Since many random factors could affect cascading failure simulation, we set  $\gamma =0.07$, $\Delta T = 11^{\circ}\mathrm{C}$ and run the model for different times in order to decide a number for which the variance of the simulation results is small enough. 
Fig. \ref{num} shows the average value and the standard deviation of the number of outages for different number of simulations. It is seen that after 10,000 simulations the average value of outages (both line outages and generator outages) stabilizes and the standard deviation of the number of outages decreases to a very small value. For the other parameters of $\gamma$ and $\Delta T$ results are very similar and are thus not given. Therefore, in this paper we run our model for 10,000 times for each case. 

\subsection{Impact of Temperature Disturbances and Size of the Selected Area}\label{temp_disturbance}

\begin{figure}[!t]
\centering
\begin{subfigure}{.24\textwidth}
  \centering
  \includegraphics[width=1.8in]{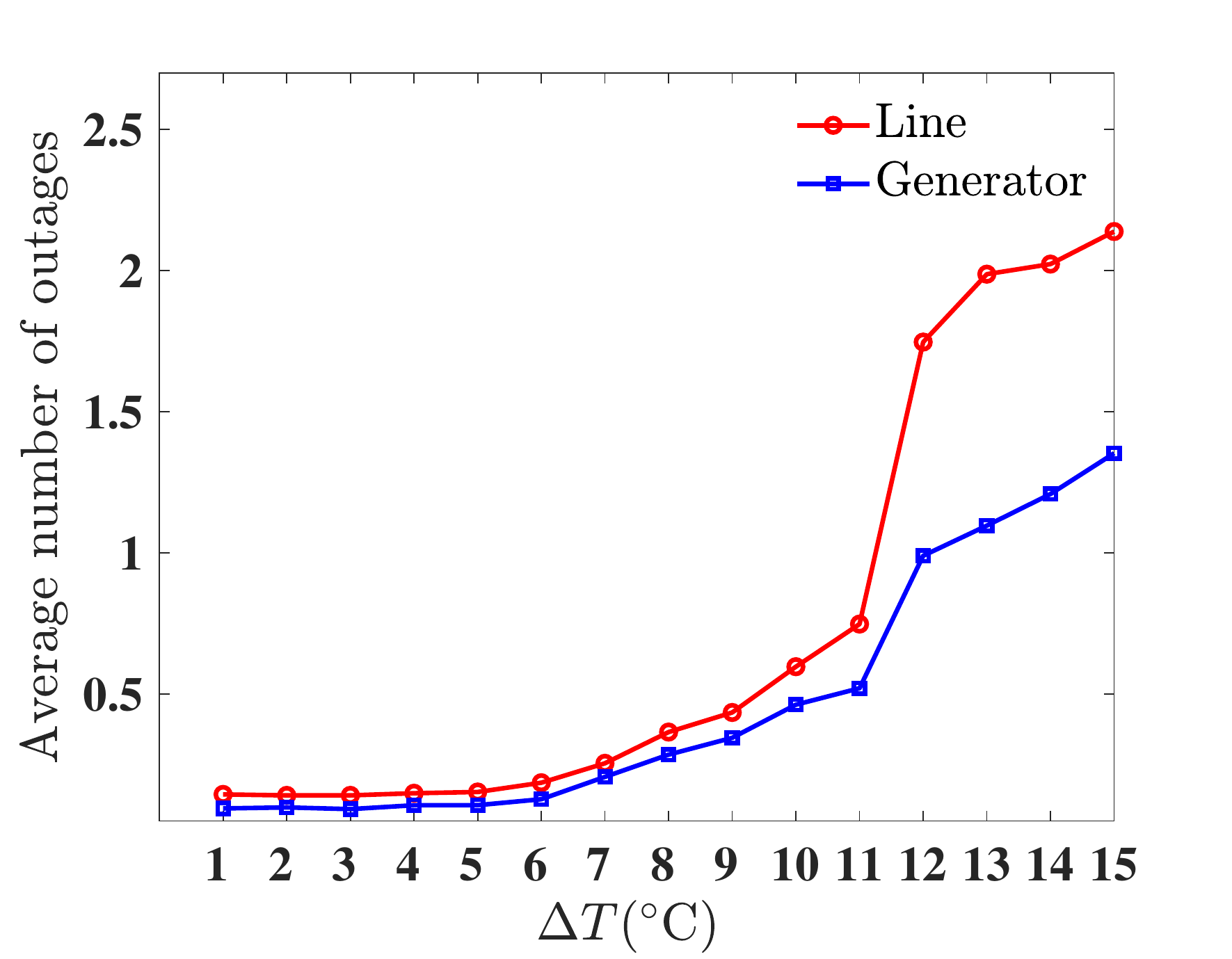}
  \caption{}
  \label{Tem_gama_fix}
\end{subfigure}%
\begin{subfigure}{.24\textwidth}
  \centering
  \includegraphics[width=1.8in]{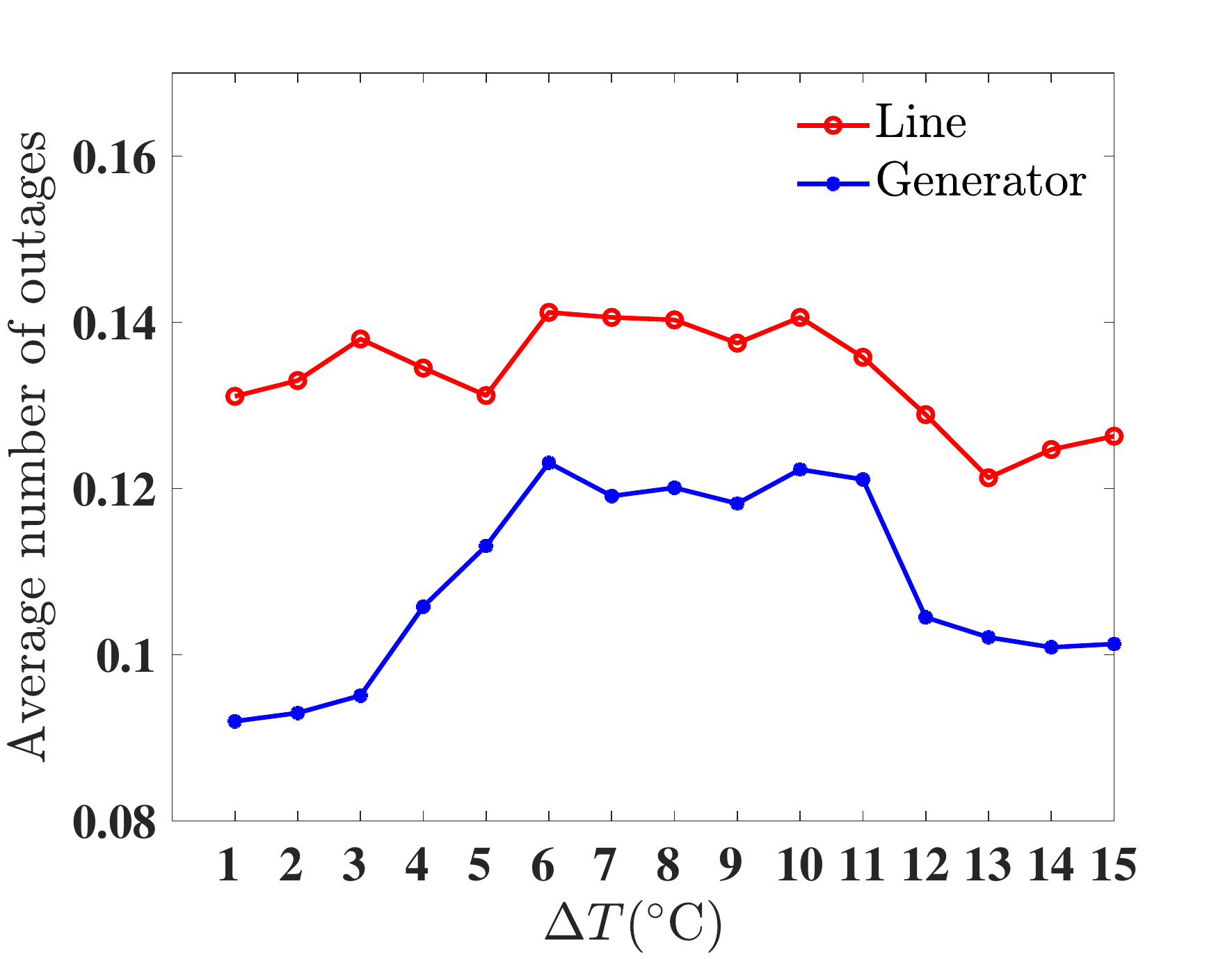}
  \caption{}
  \label{decrease}
  \end{subfigure}
  	\captionsetup{justification=raggedright,singlelinecheck=false}
  \caption{{Average value of the number of outages under different temperature disturbances: (a) Increasing temperature; (b) Decreasing temperature.} }
  \label{increase_decrease}
  \end{figure}

\begin{figure}[!t]
\centering
	\centerline{\includegraphics[width=2.0in]{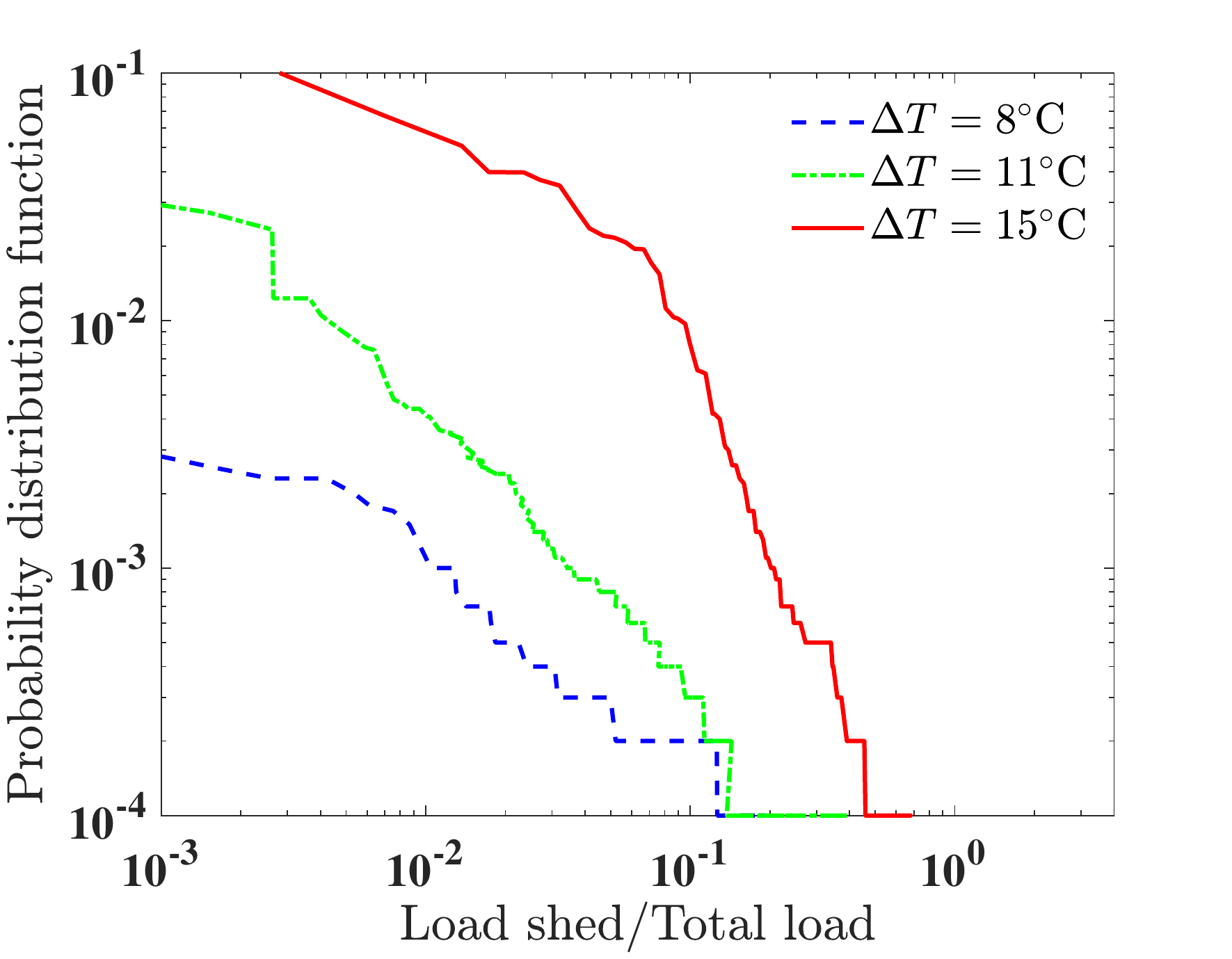}}
	\captionsetup{justification=raggedright,singlelinecheck=false}
\caption{Distribution of the load shed under different temperature increase disturbances.}
\captionsetup{justification=raggedright,singlelinecheck=false}
\label{temps_loadshed}
\end{figure}

We set $\gamma =0.07$ and run the proposed model  for 10,000 times with randomly selected areas, in which the ambient temperature is increased from $T_{\mathrm{high}}^0$ or decreased from $T_{\mathrm{low}}^0$ by $\Delta T$. Fig. \ref{increase_decrease} shows the average value of the number of outages under different temperature disturbances. It is seen that when increasing the temperature the total number of line and generator outages are low when $\Delta T \le 11^{\circ}\mathrm{C}$, and will grow quickly when $\Delta T > 11^{\circ}\mathrm{C}$. Therefore, $\Delta T=11^{\circ}\mathrm{C}$  can be inferred as the critical temperature disturbance. As presented in Fig. \ref{decrease}, if the temperature is decreased, while the load is increased the dynamic line rating is also increased. Therefore, the total number of line and generator outages is always low. Fig. \ref{temps_loadshed} shows the total load shed for three different temperature increase disturbances and it is seen that the total load shed increases under larger temperature disturbances.

\begin{figure}[!t]
\centering
	\centerline{\includegraphics[width=2.0in]{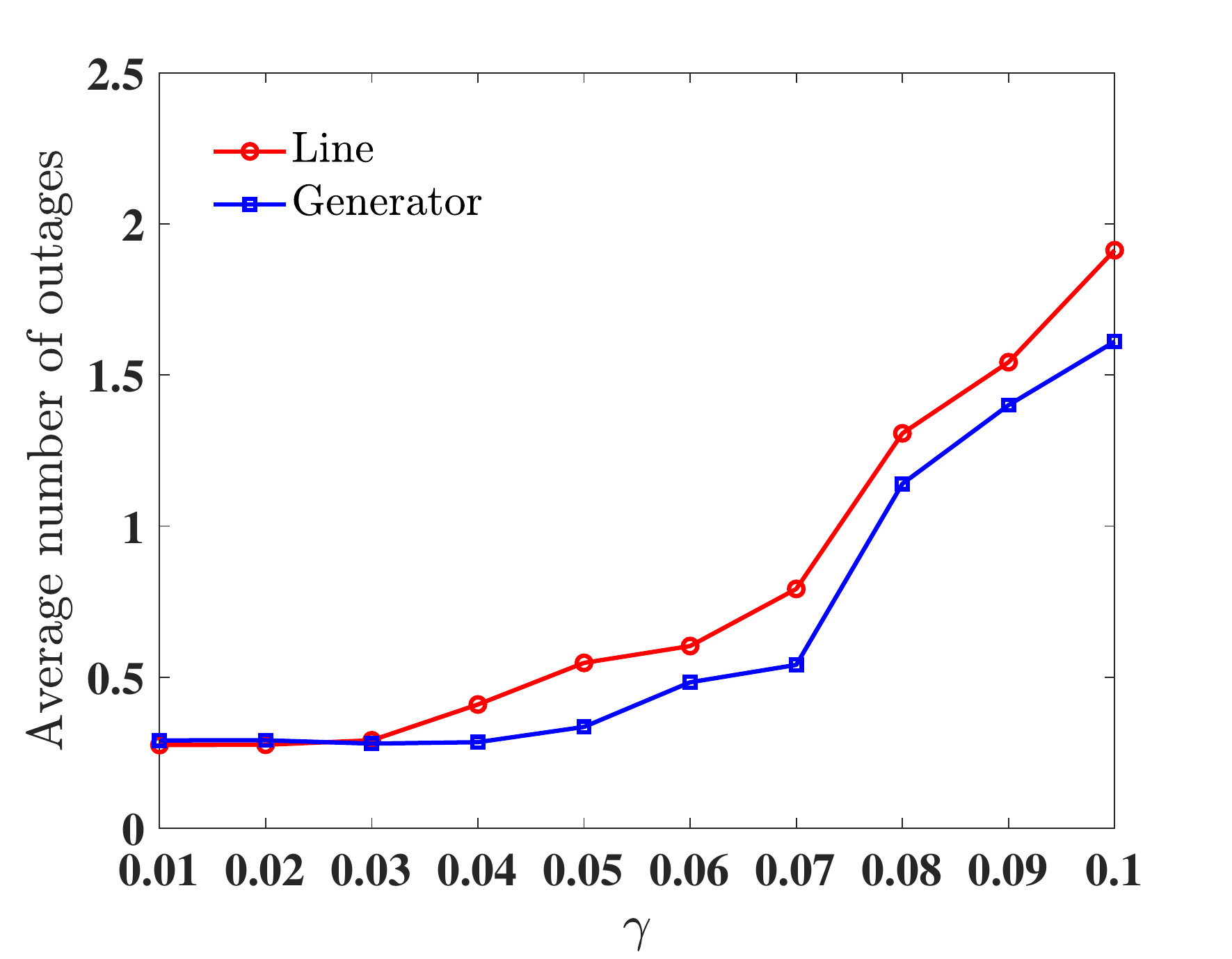}}
	\captionsetup{justification=raggedright,singlelinecheck=false}
\caption{{Average value of the number of outages under different areas.}}
\captionsetup{justification=raggedright,singlelinecheck=false}
\label{gama_temp_fix}
\end{figure}

We also analyze the impact of the size of the selected area for temperature increase cases with $\Delta T= 11^{\circ}\mathrm{C}$, and a random area is selected with a different value of $\gamma$. As the results in Fig. \ref{gama_temp_fix} indicate, by enlarging the selected area, the number of line and generator outages increases. The number of line and generator outages significantly increases for $\gamma > 0.07$. Therefore, $\gamma = 0.07$ can be inferred as the critical size.

\begin{figure}[!t]
	\centering
	\centerline{\includegraphics[width=2.0in]{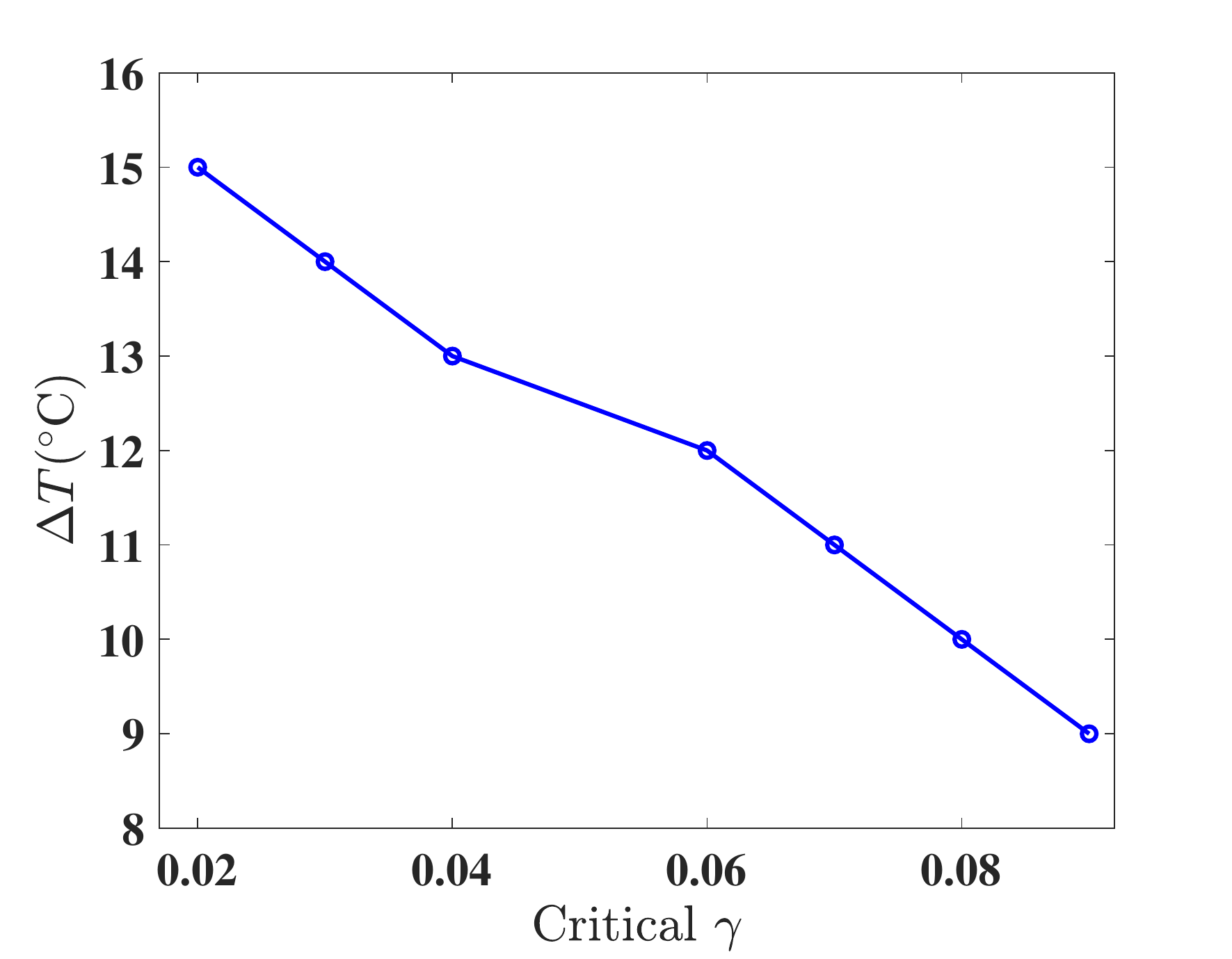}}
		\captionsetup{justification=raggedright,singlelinecheck=false}
	\caption{Critical temperature versus critical area size.}
		\captionsetup{justification=raggedright,singlelinecheck=false}
	\label{tvsgama}
\end{figure}

Fig. \ref{tvsgama} shows the pairs of critical temperature increase disturbances and critical area sizes. It can be seen that as the critical size of the selected area increases the critical temperature disturbance decreases.

\subsection{Identifying the Most Vulnerable Buses/Locations}

\begin{figure}[!t]
\centering
\begin{subfigure}{.24\textwidth}
  \centering
  \includegraphics[width=1.7in]{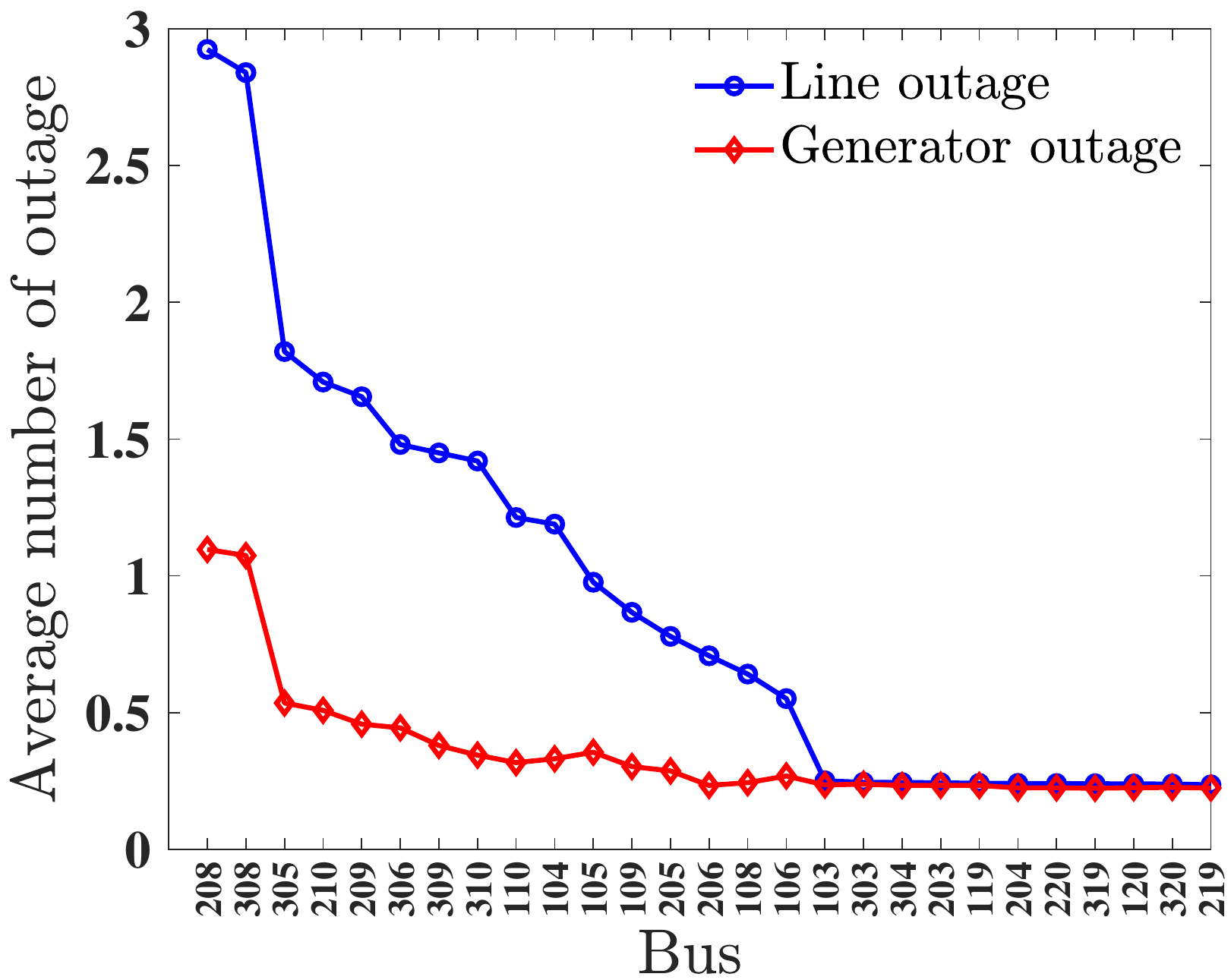}
  \caption{}
  \label{Vul}
\end{subfigure}%
\begin{subfigure}{.24\textwidth}
  \centering
  \includegraphics[width=1.7in]{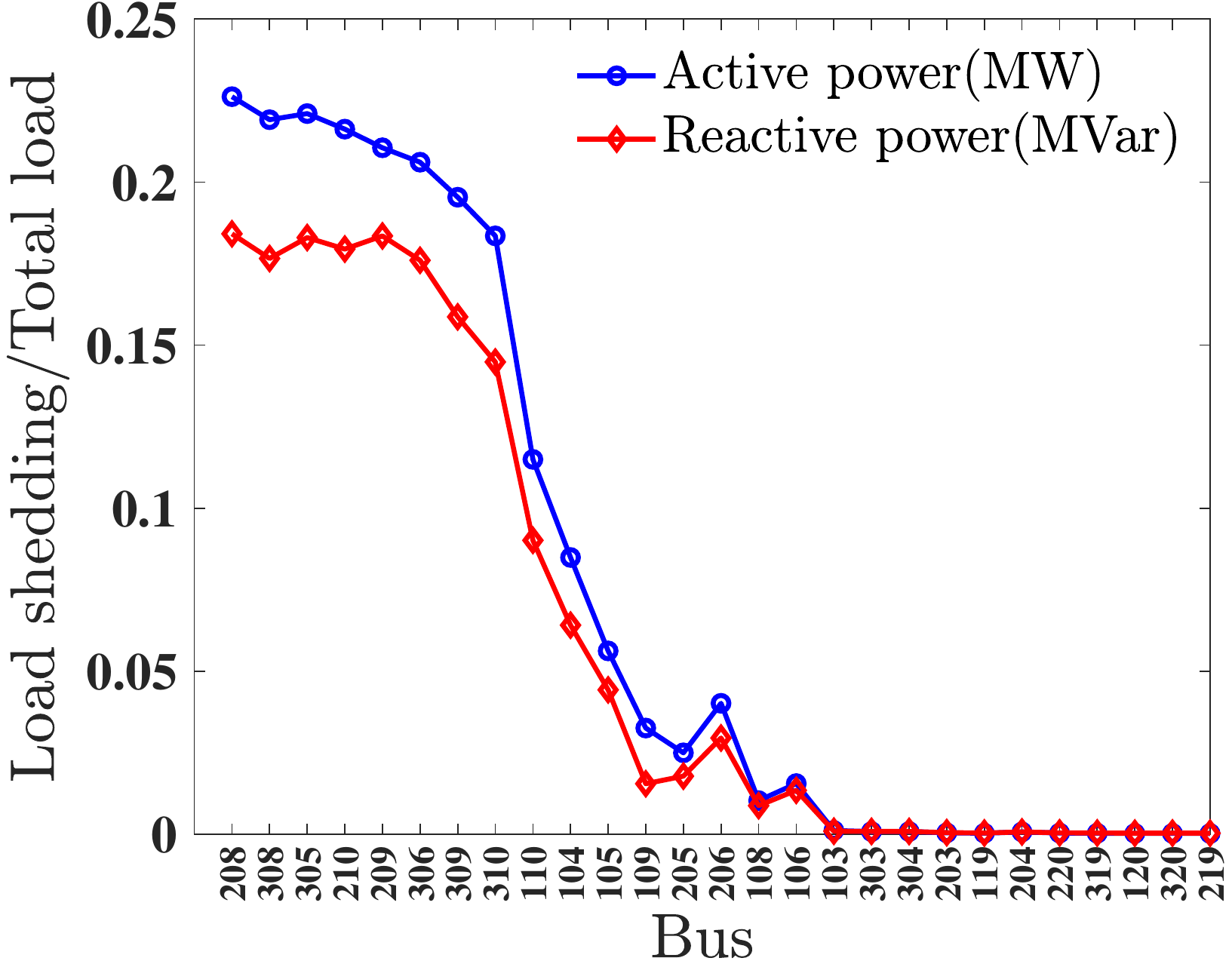}
  \caption{}
  \label{Vul_P_Q}
\end{subfigure}
\captionsetup{justification=raggedright,singlelinecheck=false}
\caption{Identification of vulnerable buses: (a) Average value of the number of outages for load buses; (b) Ratio between the load shed and the total load for load buses.}
\captionsetup{justification=raggedright,singlelinecheck=false}
\label{fig:vul}
\end{figure}

The vulnerability of the buses/locations is dependent on the temperature disturbance and size of the selected area. For this reason, for effectively identifying the most vulnerable locations, \textcolor{black}{we run the model $10000$ times for all combinations of $\Delta T = \{8 ^{\circ}\mathrm{C}, 10 ^{\circ}\mathrm{C}, 11 ^{\circ}\mathrm{C}, 15 ^{\circ}\mathrm{C}\}$ and $\gamma = \{0.05, 0.06, 0.07, 0.08\}$  around every load bus. By doing so we can identify the vulnerable buses under very diverse failure scenarios. } Fig. \ref{fig:vul} shows the simulation results. Fig. \ref{Vul} shows the average value of the number of line and generator outages and Fig. \ref{Vul_P_Q} shows the average value of the ratio between the load shed and the total load. We can see that the load buses 208, 308, 305, 210, 209, 306, 309, and 310 are more vulnerable than the other buses, and the temperature disturbances around these buses lead to much more line/generator outages and load shed.

\subsection{Impact of Control Strategies}

For the typical case presented in Section \ref{typical_case}, if the operator re-dispatch is modeled by considering the initial line rating, at 8972 s the operator re-dispatch is performed because the power flow of some branches exceeds their initial capacities and the  potential outage of line 209-211 would take 104.4 s which is longer than 60 s, the time required for operator re-dispatch. After the operator re-dispatch the cascading failure stops.  

\begin{figure}[htb!]
\centering
\begin{subfigure}{.24\textwidth}
  \centering
  \includegraphics[width=1.8in]{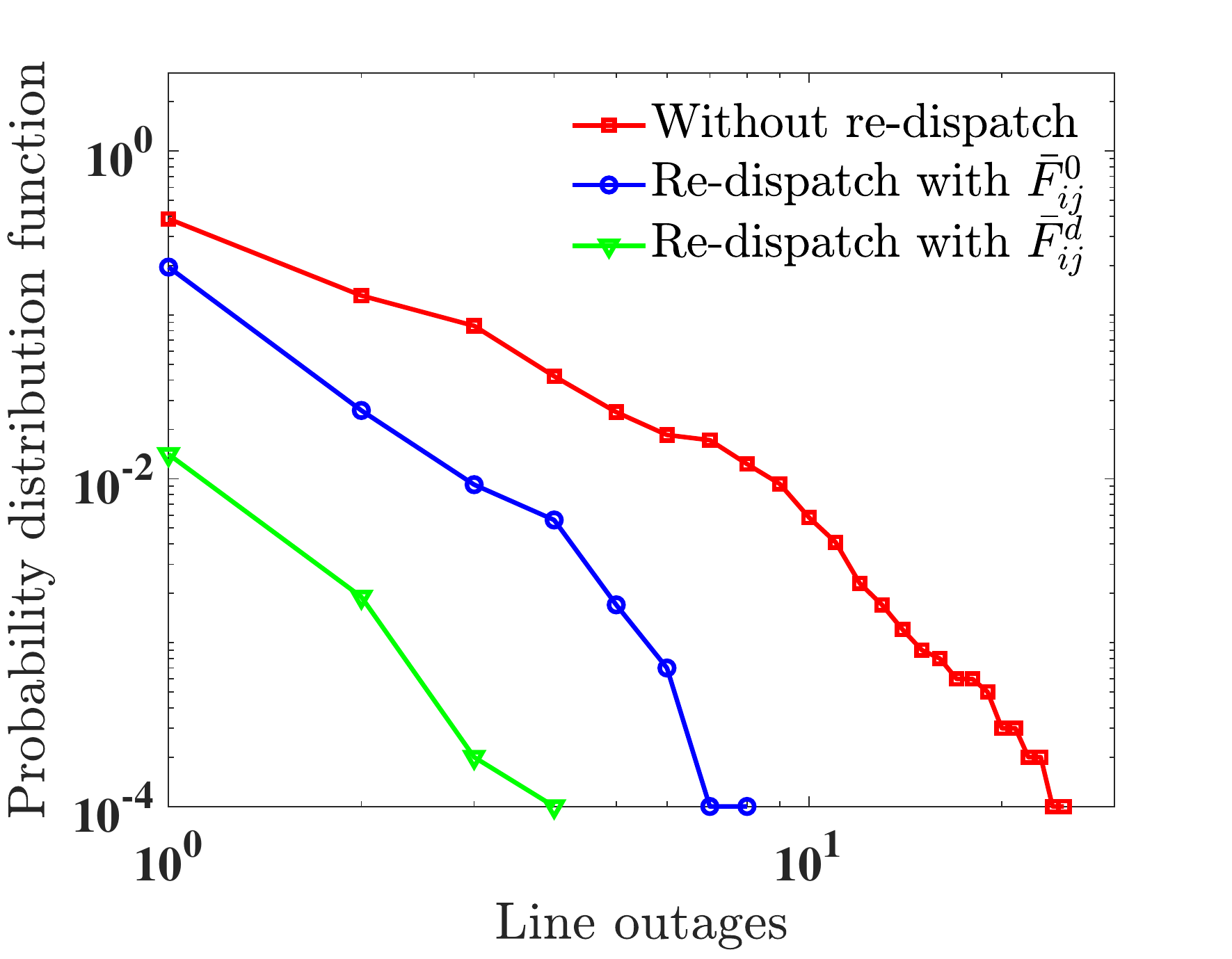}
  \caption{}
  \label{L_four}
\end{subfigure}%
\begin{subfigure}{.24\textwidth}
  \centering
  \includegraphics[width=1.8in]{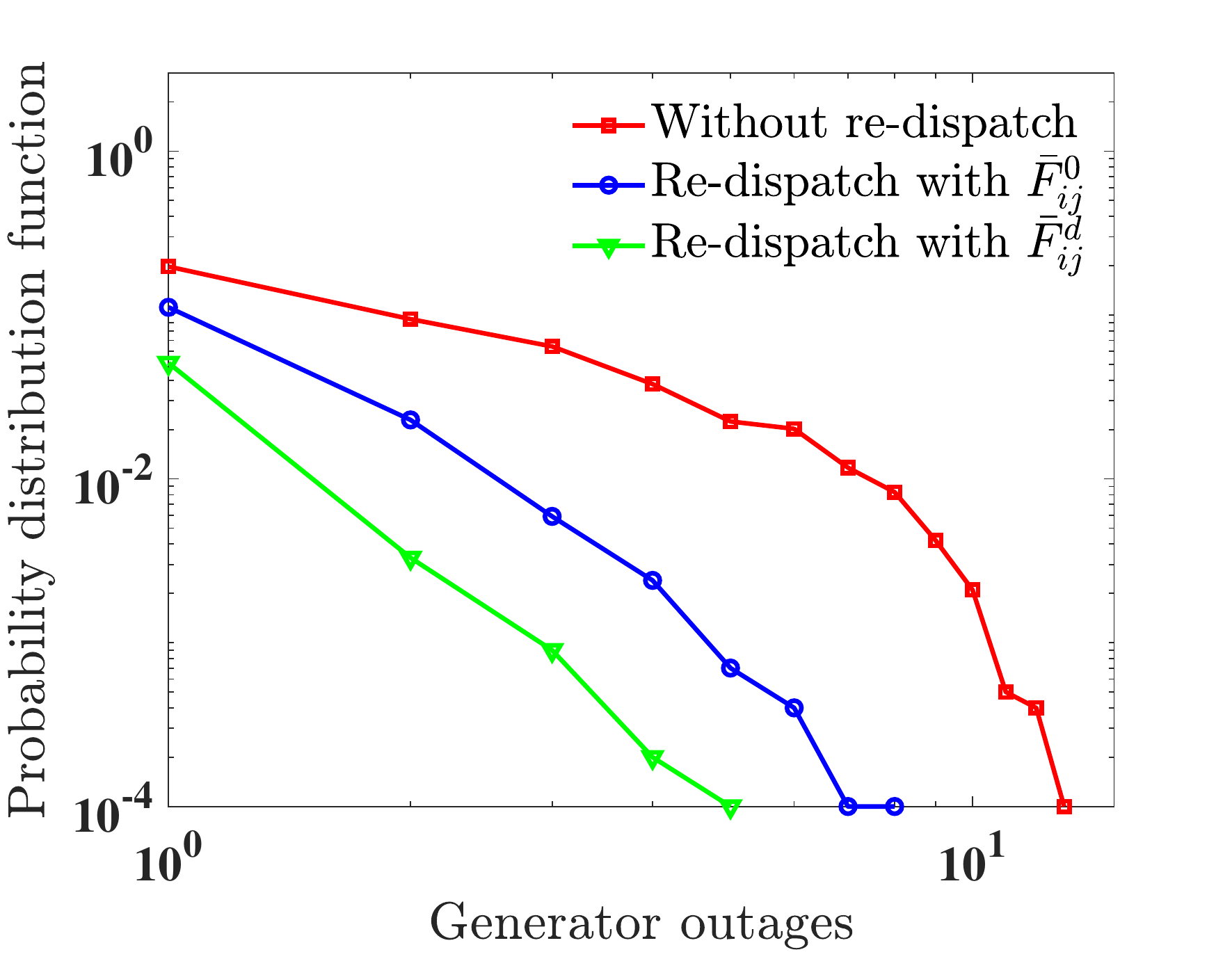}
  \caption{}
  \label{G_four}
\end{subfigure}
\captionsetup{justification=raggedright,singlelinecheck=false}
\caption{Distribution of the number of line and generator outages with or without re-dispatch. (a) Line outage (b) Generator outage.}
\label{fig:lgthree}
\end{figure}

Besides, we set $\Delta T= 11^{\circ}\mathrm{C}$, $\gamma=0.07$ and run simulations for 10,000 times for the cases with or without considering operator re-dispatch.  Fig. \ref{fig:lgthree} shows the distributions of the number of line and generator outages. It is clear that by implementing the re-dispatch strategy using $F_{ij}^0$, the number of line and generator outages is decreased. Besides, by considering dynamic line rating for re-dispatch, the risk can be reduced much more significantly.

\section{Conclusion} \label{conclusion}

In this paper, a blackout model that considers load change and dynamic line rate change due to ambient temperature disturbance is proposed. We apply the proposed model to the RTS-96 3-area system and find that temperature disturbance can lead to correlated load change and line tripping, which will together contribute to voltage collapse. Based on the developed model we identified critical temperature change, critical area with temperature disturbance, and most vulnerable buses, and compare the effectiveness of different control strategies. 

In this paper the major mechanisms that could lead to the initiating and propagation of cascading failures are the load increase and line rating decrease caused by ambient temperature disturbances, the coupling between different events such as line outages, generator outages, and undervoltage of load buses, and also the consequent voltage collapse. As the penetration of renewable generation is quickly increasing, the future power system will be even more impacted by external factors such as weather conditions. This is because compared with traditional fossil fuel based generation the renewable, mostly power electronics interfaced generation depends more on weather conditions, and is more sensitive to the system disturbances such as voltage disturbances caused by transmission outages due to lightning or wildfire \cite{sc_disturbance,yan2018anatomy,UKBLACKOUT}. The blackout model proposed in this paper can be further extended for the future power system with high penetration of renewable generation to evaluate the cascading failure risk, identify critical components that play important roles in outage initiating and propagation, and develop effective mitigation strategies to significantly reduce the cascading risk.

\bibliographystyle{IEEEtran}
\bibliography{final}

\footnotesize \textbf{Seyyed Rashid Khazeiynasab} received the B.Sc. degree from Shiraz University,
Shiraz, Iran, and the M.Sc. degree from
Sharif University of Technology, Tehran, Iran, in
2011 and 2013 respectively. He joined University
of Central Florida, USA in August 2018.
Currently, he is pursuing his Ph.D. in Electrical Engineering
under the supervision of Dr. Junjian Qi. His research interests include cascading blackouts and parameter estimation.

\vspace{0.1in}
\footnotesize \textbf{Junjian Qi} received the B.E. degree in electrical engineering, from Shandong University, Jinan, China, in 2008, and the Ph.D. degree in electrical engineering from Tsinghua University, Beijing, China, in 2013. He was a Visiting Scholar with Iowa State University, Ames, IA, USA, in 2012, a Research Associate with the Department of EECS, University of Tennessee, Knoxville, TN, USA, from 2013 to 2015, a Postdoctoral Appointee with the Energy Systems Division, Argonne National Laboratory, Lemont, IL, USA, from 2015 to 2017, and an Assistant Professor with the Department of Electrical and Computer Engineering, University of Central Florida, Orlando, FL, USA, from 2017 to 2020. He is currently an Assistant Professor with the Department of Electrical and Computer Engineering, Stevens Institute of Technology, Hoboken, NJ, USA. He was the recipient of the NSF CAREER award in 2020 and is an Associate Editor for the IEEE Access. His research interests include cascading blackouts, microgrid control, cyber-physical systems, and synchrophasors. 

\end{document}